\documentclass[conference]{IEEEtran}
\usepackage[sorting=none]{biblatex}
\usepackage{url}
\usepackage{hyperref}

\usepackage{array}
\newcolumntype{P}[1]{>{\centering\arraybackslash}p{#1}}
\newcolumntype{L}[1]{>{\centering\arraybackslash}l{#1}}
\usepackage{threeparttable}

\ifCLASSINFOpdf
  \usepackage[pdftex]{graphicx}
  \graphicspath{{../pdf/}{../jpeg/}}
  \DeclareGraphicsExtensions{.pdf,.jpeg,.png}
\else
\fi
\usepackage{amsmath}
\usepackage{bm}
\def\*#1{\mathbf{#1}}
\def\'#1{\boldsymbol{#1}}

\usepackage{stfloats}
\usepackage{subcaption}
\usepackage{amssymb}
\usepackage{algpseudocode}

\usepackage[linesnumbered,ruled,vlined]{algorithm2e}
\usepackage{cleveref}
\crefname{algorithm}{algorithm}{algorithms}
\Crefname{algorithm}{Algorithm}{Algorithms}

\usepackage{multirow}
\usepackage[table]{xcolor}
\usepackage{array}
\usepackage{threeparttable}

\usepackage{enumitem}

\begin{document}

\title{Condition Monitoring with Machine Learning: A Data-Driven Framework for Quantifying Wind Turbine Energy Loss}
\thispagestyle{plain}
\pagestyle{plain}

\author{\IEEEauthorblockN{Emil~Marcus~Buchberg}
\IEEEauthorblockA{\textit{Institute of Computer Science} \\
\textit{Aalborg University}\\
Aalborg, Denmark \\
ebuchb20@student.aau.dk}
\and
\IEEEauthorblockN{Kent~Vugs~Nielsen}
\IEEEauthorblockA{\textit{Institute of Computer Science} \\
\textit{Aalborg University}\\
Aalborg, Denmark \\
kniels18@student.aau.dk}
}

\maketitle


\begin{abstract}
Wind energy significantly contributes to the global shift towards renewable energy, yet operational challenges, such as Leading-Edge Erosion on wind turbine blades, notably reduce energy output. This study introduces an advanced, scalable machine learning framework for condition monitoring of wind turbines, specifically targeting improved detection of anomalies using Supervisory Control and Data Acquisition data. The framework effectively isolates normal turbine behavior through rigorous preprocessing, incorporating domain-specific rules and anomaly detection filters, including Gaussian Mixture Models and a predictive power score. The data cleaning and feature selection process enables identification of deviations indicative of performance degradation, facilitating estimates of annual energy production losses. The data preprocessing methods resulted in significant data reduction, retaining on average 31\% of the original SCADA data per wind farm. Notably, 24 out of 35 turbines exhibited clear performance declines. At the same time, seven improved, and four showed no significant changes when employing the power curve feature set, which consisted of wind speed and ambient temperature. Models such as Random Forest, XGBoost, and KNN consistently captured subtle but persistent declines in turbine performance. The developed framework provides a novel approach to existing condition monitoring methodologies by isolating normal operational data and estimating annual energy loss, which can be a key part in reducing maintenance expenditures and mitigating economic impacts from turbine downtime.

\end{abstract}
\section{Introduction}\label{sec:introduction}

\IEEEPARstart{W}{ind} energy plays a critical role in the global shift towards renewable energy sources, prominently contributing to meeting targets such as the European Union’s climate neutrality goal by 2050. Wind turbines (WTs) have proven to be a reliable renewable energy source, increasingly affordable compared to other alternatives \cite{EUWind}. However, WTs encounter notable production and operational challenges due to the inherently stochastic nature of their operational environments, necessitating regular maintenance and repair activities. These activities are essential to maintain reliability, such as consistent energy production, continuous availability for operational uptime, and safe operation of turbines. Economic losses frequently arise from such maintenance and repair activities, often due to downtime resulting in halted turbine operations. Consequently, there is significant interest in minimizing downtime, particularly in the current context where energy prices decrease. Still, operation and maintenance costs of wind farms (WFs) have remained relatively constant over the past decade, making up one-quarter of the lifetime cost of onshore farms and one-third of the offshore farm cost \cite{IRENA}. Despite the comparatively high operation and maintenance costs associated with offshore wind farms, these installations continue to draw increased attention to their locational flexibility and the consistently higher wind speeds available at sea. Unlike many onshore installations, they are largely unconstrained by residential noise restrictions, permitting the deployment of larger blade dimensions and higher tip speeds, enhancing energy yield. Nevertheless, the offshore setting imposes considerable logistical challenges and elevates operation and maintenance costs \cite{papatheou2014wind}\cite{chesterman2021condition}\cite{oliveira2024wind}\cite{izquierdo2019framework}. Due to higher blade velocities and increased exposure to environmental factors such as wind, rain droplets, hailstones, sand grains, frost, sea spray, temperature oscillations, ultraviolet radiation, and fog, all of which accelerate material degradation \cite{cappugi2021machine}\cite{duthe2021modeling}\cite{visbech2023introducing}\cite{castorrini2024opensource}. The damage to the blade airfoils alters their geometry, thereby degrading aerodynamic performance and resulting in increased drag and decreased lift forces, thereby diminishing overall turbine performance. This phenomenon is referred to as Leading-Edge Erosion (LEE) and is characterized by a gradual manifestation of physiological degradation on WT blades, leading to a progressive decline in energy production \cite{cappugi2021machine}\cite{ehrmann2017effect}\cite{sareen2014effects}\cite{bak2020influence}\cite{kruse2018predicting}\cite{schramm2017influence}. Although onshore turbines are subject to similar challenges, the incidence and severity of LEE are more predominant in offshore environments and highlights the importance of accounting for these effects in subsequent analyses.

To mitigate these economic impacts, Condition Monitoring (CM) strategies are advocated, aimed at the early detection of degradation and isolation of incipient faults. CM facilitates condition-based maintenance strategies capable of outperforming traditional scheduled maintenance by identifying component degradation early and minimizing unnecessary downtime. Typically, CM strategies leverage sensor networks integrated within Supervisory Control and Data Acquisition (SCADA) systems, which continuously collect operational data, enabling predictive maintenance approaches \cite{papatheou2014wind}\cite{chesterman2021condition}\cite{oliveira2024wind}\cite{izquierdo2019framework}\cite{bonacina2022use}\cite{meyer2021multi}\cite{wang2014copula}\cite{yasuda2017system}\cite{bilendo2021intelligent}\cite{jin2022condition}\cite{meyer2020data}. A key monitoring tool in the context of CM is the wind turbine power curve (PC), an essential characterization linking wind speed with turbine power output. SCADA systems routinely record operational and meteorological data, typically at 10-minute averaging intervals, which are instrumental in constructing and monitoring the PC. However, SCADA systems face limitations such as sensor errors, data loss due to communication errors, and other data-quality issues such as spurious entries arising from the averaging process \cite{wang2014copula}. These spurious entries can be a result of various operational states, such as shifts between power-production states, idling, and curtailment, which can drive large signal fluctuations that obscure fault signatures.

Machine Learning (ML)-based methodologies have increasingly gained traction in CM practices and are employed in real-time applications to detect deviations indicative of anomalies or incipient failures. However, supervised ML models face a significant challenge of strong class imbalance arising from rare fault occurrences and difficulties associated with labeling high-dimensional SCADA data. Thus, reference models are often established using historical data from healthy operational states. The reason for training models on healthy operational states, also known as normal behavior (NB) data, is that it is often more feasible to learn a representation of the turbine’s NB and detect relevant deviations from this behavior. Defining NB remains a major challenge in itself because it may change over time, also known as concept drift, owing to events such as software updates, sensor recalibrations, part replacements, or slow processes associated with normal aging such as LEE. \cite{meyer2021multi}. Thus, integrating complex CM strategies with ML frameworks and well-defined NB data remains essential to optimizing wind farm performance, minimizing production losses, and reducing maintenance expenditures \cite{bonacina2022use}.

The body of research on data-driven techniques for monitoring WT health has expanded noticeably, yet important methodological shortcomings remain unaddressed. Some studies identify outliers solely through visual inspection of the PC, a process that is neither objective nor reproducible \cite{wang2014copula}. Only a small fraction of existing studies engages explicitly with abnormal or erroneous SCADA data. Many either ignore such records or remove them without systematic investigation. Even among works that attempt anomaly detection, most rely on relatively simple unsupervised algorithms and single-signal models, unsuited to the high dimensionality and complex operating modes characteristic of turbine SCADA data \cite{meyer2021multi}. For multiple WTs across multiple WFs, this one-model-per-signal paradigm quickly scales to a plethora of models and threshold values, imposing a maintenance burden that has scarcely been discussed in the literature \cite{meyer2021multi}. Others proceed directly to anomaly detection without first removing obvious outliers, which can lead to datasets where anomalies distort the statistical notion of normality due to stacked data points. Distance and density-based methods such as Local Outlier Factor or DBSCAN struggle in these circumstances \cite{morrison2022anomaly}. As a result, the difficulty of isolating truly normal behavior data for model training and interpreting prediction errors is still widely acknowledged but rarely resolved. Among the papers referenced, only \cite{morrison2022anomaly} proposes a comprehensive framework that confronts most of the complexity in WT SCADA data. \cite{morrison2022anomaly} also voices concerns about the field’s limited progress on these issues. We share this assessment and emphasize that treatment of abnormal data, scalable multi-target modeling strategies, and systematic pre-processing pipelines are open problems that must be addressed to advance CM of WTs.

\medskip

This project builds on our earlier research, conducted in collaboration with the Danish energy company Vestas, on LEE and its impact on WT performance \cite{VUGS_AND_BUCH_TO_THE_TOP}. Our previous study developed an ML pipeline that utilized offshore SCADA data to detect LEE-induced deviations and to translate them into estimates of lost energy production. While the framework demonstrated the feasibility of data-driven LEE detection, it revealed several unresolved challenges:

\begin{itemize}
    \item Can identifying and modeling healthy operating states be improved in an unsupervised setting.
    \item Can a scalable approach accommodate the high dimensionality of SCADA streams across large WFs.
    \item Can deviations from normal behavior be converted into more reliable estimates of energy production loss.
\end{itemize}

Addressing these questions will advance a data-driven methodology capable of detecting anomalous operations while providing actionable estimates of production losses and degradation rates for WTs. The present study refines and generalizes the previous framework to support scalable predictive maintenance strategies for large-scale WFs. Central to this approach are normal behavior filters (NB-filters), which employ statistical and ML techniques to isolate operational data representative of normal performance. To complement the NB-filters, we introduce hard-filters, a more stringent set of rules designed to remove obvious outliers before NB-filtering. The formulation and implementation of both filter classes are detailed in \Cref{sec:normality}. Additionally, we include a Predictive Power Score (PPS) that captures the relationship among key SCADA variables. The combined average PPS provides an indicator of each turbine’s operational state.

\medskip

The remainder of the paper is organized as follows. \Cref{sec:related} references data-driven approaches to WT condition monitoring, highlighting how our methodology departs from and advances state-of-the-art. \Cref{sec:dataset} describes the SCADA data set, its sampling characteristics, and the raw-data quality issues, including representative outliers, that motivate subsequent filtering. \Cref{sec:preliminary_concepts} introduces notation and preliminary concepts needed to reason with the subsequent sections. \Cref{sec:normality} formalizes the concept of normal turbine operation, introduces the hard-filter and NB-filter pipelines, and justifies their use. \Cref{sec:ml-strategies} details the learning algorithms and the training and validation strategies. \Cref{sec:experiments} specifies the hyper-parameter search spaces, outlines two distinct experiments, and presents the evaluation metrics. \Cref{sec:results} reports the impact of hard-filters and NB-filters and presents our findings from the two experiments included. \Cref{sec:discussion} interprets the results in light of practical deployment, theoretical expectations, and limitations. \Cref{sec:conlusion} synthesises the principal contributions, while \Cref{sec:future-work} proposes avenues for extending the framework.
\section{Related work}\label{sec:related}
Numerous studies have attempted to quantify the long-term drift accumulating in the energy production efficiency of WTs with a diverse range of propositions, including our own \cite{VUGS_AND_BUCH_TO_THE_TOP}. Due to the diverse applicable methodologies in the research domain, the reported effect of LEE on annual energy production (AEP) varies significantly. In \cite{VUGS_AND_BUCH_TO_THE_TOP}, we specifically find the worst-case scenario
to be a 2.12\% energy production loss over 3 years, and present findings from other studies, reporting the AEP loss to range from $1\%$ to $25\%$, depending on the study and the severity of the accumulated LEE, signaling the need for a unified and precise approach of quantification \cite{cappugi2021machine}\cite{ehrmann2017effect}\cite{sareen2014effects}\cite{bak2020influence}\cite{kruse2018predicting}\cite{schramm2017influence}\cite{panthi2023quantification}. To assess the effects of LEE, a theoretical model can often be established similar to \cite{duthe2021modeling}\cite{bak2020influence}\cite{ozccakmak2024determination}, and the AEP loss or LEE can be measured through simulations or observations in deviations from the established model. These types of models are often constructed under controlled environments with \cite{ehrmann2017effect} and \cite{schramm2017influence} being simulation-based, \cite{sareen2014effects} being tested in a wind tunnel, and \cite{visbech2023introducing} being partially simulated, using a combination of observed erosion with synthetically generated weather data. These theoretical models and studies play a critical role in understanding the direct impact of LEE and AEP loss and provide fundamental principles applicable to practical modeling. However, as argued by \cite{pandit2022scada}, theoretical models cannot always reflect the site and turbine-specific operational behavior. A WT is exposed to site-specific weather conditions, wake effects, sensor and component degradations, and other complex WT dynamics. This claim is further supported by \cite{meyer2021multi}\cite{pandit2022scada}, maintaining that either WT or site-specific behavior should be learned under a data-driven framework directly from the operation history. Both the theoretical models and the data-driven framework often fall under the category of NB models and frequently serve as a point of reference for further inference. This study focuses mainly on the group of data-driven frameworks and the models derived from such frameworks. The consensus in the literature is that the effect of LEE on AEP reduces the power output year-on-year. However, for much of the NB-related research, the effects of LEE directly on AEP are not measurable, and the slow deviation from established NB models is built on the assumption that LEE is the main causation. In this study, we similarly assume that a slow decline in performance is caused by latent variables such as LEE.
\begin{table*}[t]
    \centering
    \caption{WF Data Points Overview}
    \label{tab:turbine_data}
    \begin{tabular}{|c|c|c|c|c|c|c|c|}
        \hline
        \textbf{WF Id} & \textbf{Location} & \textbf{Start Year} & \textbf{End Year} & \textbf{Turbine Ids} & \textbf{Data Points} & \textbf{Avg. Data Points per Turbine} \\ \hline
        1 & Offshore & 2017 & 2024 & 1-10 & 3,993,086 & 399,308.60 \\ \hline
        2 & Offshore & 2017 & 2024 & 11-20 & 3,607,779 & 360,777.90 \\ \hline
        3 & Offshore & 2017 & 2024 & 21-25 & 1,742,279 & 348,455.80 \\ \hline
        5 & Offshore & 2017 & 2024 & 26-35 & 2,715,567 & 271,556.70 \\ \hline
        6 & Onshore & 2015 & 2024 & 36-40 & 2,557,851 & 511,570.20 \\ \hline
        7 & Onshore & 2015 & 2024 & 41-45 & 2,323,400 & 464,680.00 \\ \hline
        8 & Onshore & 2015 & 2024 & 46-54 & 4,427,375 & 491,930.56 \\ \hline
        9 & Onshore & 2015 & 2024 & 55-60 & 3,025,880 & 504,313.33 \\ \hline
        10 & Onshore & 2015 & 2024 & 61-67 & 3,549,919 & 507,131.29 \\ \hline
        11 & Onshore & 2015 & 2024 & 68-76 & 4,135,727 & 459,525.22 \\ \hline
        12 & Onshore & 2015 & 2024 & 77-83 & 3,203,512 & 457,644.57 \\ \hline
        13 & Onshore & 2015 & 2024 & 84-90 & 3,551,612 & 507,373.14 \\ \hline
        14 & Onshore & 2015 & 2024 & 91-99 & 3,661,406 & 406,822.89 \\ \hline
        15 & Onshore & 2015 & 2024 & 100-108 & 4,598,804 & 510,978.22 \\ \hline
        16 & Onshore & 2015 & 2024 & 109-117 & 4,579,266 & 508,807.33 \\ \hline
    \end{tabular}
\end{table*}
Loosely defined for this section and briefly hinted in \Cref{sec:introduction}; an NB model is a data-driven representation of how a system is expected to operate under fault-free or healthy conditions. ML NB models are trained on healthy periods of operation data to reflect a healthy state reference point or period of operation history that can be used directly for prediction and inference \cite{barnabei2024nbm}\cite{bilendo2023applications}. In a continuation of \cite{meyer2020data}, Meyer in  \cite{meyer2021multi} trains several multi-target NB ML regression models to predict normal turbine behavior learned from the historical SCADA data and highlights the precision and potential of ML regressors in NB modeling. The CM strategy employed in both studies is performed as a residual analysis, meaning high deviations between the measured observations and the NB model predictions will produce statistical changes in the residual distribution and be detected when predefined thresholds are exceeded. The models showcase low prediction errors and suggest a potential for identifying turbine underperformance by utilizing the predictions generated by the NB models. Building further on this idea are studies such as \cite{Byrne2020}, \cite{Astolfi2021V52Aging}, \cite{Astolfi2022}, and \cite{mathew2022estimation} that depart from the CM strategy of NB models and instead focus on quantifying the long-term drift potentially found in the predictions. Although not explicitly stating the use of NB models, their methods align with the general agreement of establishing NB models for creating system behavioral reference points. \cite{mathew2022estimation} trains a benchmark deep neural network NB model on historical operational data and predicts the subsequent operating history. The predictions reflect the expected WT behavior, which can be directly compared to the observed power output. They calculate an efficiency index $( \eta_I)$;
\begin{equation*}
    \eta_I=\dfrac{\eta_{T_{Measured}}}{\eta_{T_{Modelled}}},
\end{equation*}
with $\eta_{T_{Measured}}$ being the measured power production efficiency of the turbine and $\eta_{T_{Modelled}}$ being the corresponding period predicted by the model. The $\eta_I$ is evaluated through a time series with yearly aggregates, and they report a performance declining trend over time. \cite{Byrne2020}, \cite{Astolfi2021V52Aging}, and \cite{Astolfi2022} employ a similar data-driven approach, splitting the data into a target dataset $D_2$, a reference dataset $D_1$ and a training dataset $D_0$. $D_0$ is used for training validation and testing for a support vector regression model, while the subsequent data sets are used for prediction. A drift score $\Delta_i$ for every $D_i$, $i>0$, is then calculated, and a drift quantification for every target year where $i>1$ can be computed as $\delta_{i}=\Delta_i-\Delta_{i-j}$, where $i>j>0$. For the worst case scenario, using this method, \cite{Astolfi2021V52Aging} reports a performance degradation of up to $8.8\%$ over a 12-year period.

A common weakness among these studies is the dependency on long historical periods of healthy data, as opposed to older methods like PC binning and similar \cite{bilendo2023applications}. However, as argued by \cite{bilendo2023applications}, less sophisticated methods like binning often trade information for simplicity. In a domain where the pursued quantifications are often smaller than the model inaccuracies, researchers recently applied more complex models like the neural network of \cite{mathew2022estimation}. By implementing methods of \cite{Byrne2020}, \cite{Astolfi2021V52Aging}, \cite{Astolfi2022}, and \cite{mathew2022estimation} the reliance on healthy and operational SCADA data is evident. Plenty of research share the sentiment of healthy turbine training data requirement like \cite{bilendo2023applications} and \cite{chesterman2023overview}   However, \cite{barnabei2024nbm} argues a deep understanding of specific WT is required to correctly isolate an operating region and period in time, to derive a reliable NB model. A viewpoint we partly report in \cite{VUGS_AND_BUCH_TO_THE_TOP}, where it is demonstrated how selecting the correct training year is critical for the success of establishing a stable NB model. A promising CM method to engage with the complexity of NB data selection is used by \cite{bonacina2022use} and \cite{barnabei2024nbm}, introduced as the Combined Predictive Power Score in \cite{Miele2022}. By augmenting the original PPS \cite{wetschoreck_krabel_krishnamurthy_2020} to operate within temporal context, \cite{bonacina2022use} showcase how the Combined Predictive Power Score can be utilized to identify stable NB periods. 

The SCADA data available in this study, along with \cite{VUGS_AND_BUCH_TO_THE_TOP}, do not include event logs and similar for anonymization reasons (see \Cref{sec:dataset}), making it a non-trivial task to select stable NB data used for training. We report in \cite{VUGS_AND_BUCH_TO_THE_TOP} how WTs often experience upstart calibration, re-calibration periods, and repairs throughout installation and operational history, requiring clearly defined criteria for NB data selection. As mentioned in \Cref{sec:introduction}, appropriately selecting operation states and isolating NB data remains widely unaddressed in the NB research field. For example, in studies like \cite{meyer2021multi} the exact data selection criteria are largely unknown, with no clear examples of data cleaning addressed. \cite{Byrne2020} does not address any particular data cleaning methods, and \cite{Astolfi2022} requests data with full uptime and only filters by operation regions. \cite{Astolfi2021V52Aging} reports that they are using outlier analysis, with no further details added. In \cite{VUGS_AND_BUCH_TO_THE_TOP}, we applied rule-based filters and experimented with the Local Outlier Factor, a commonly used algorithm in the NB modeling field. However, the algorithm displayed unsatisfactory results, akin to the issues reported by \cite{morrison2022anomaly}, with stacked outliers densities too severe to deal with.

The intended outcome of this research is to address the unstable NB modeling we did in \cite{VUGS_AND_BUCH_TO_THE_TOP} and train more stable and precise models for better quantification of AEP loss. Operating within that perspective, we contextualize our methods with the general critique from \cite{morrison2022anomaly}, what we learned from \cite{VUGS_AND_BUCH_TO_THE_TOP}, and correspondence with Vestas. We engage with the general inadequate data cleaning in \cite{VUGS_AND_BUCH_TO_THE_TOP} and other similar research by building several layers of rule-based, density-based, and voting-based methods that aim to better a time-dependent PPS (similar to \cite{bonacina2022use}), to ensure the learned NB is as representative as possible. To our knowledge, very few studies engage in such a thorough data-cleaning process. Furthermore, we transparently define and select NB periods based on the PPS to ensure we automatically process WTs with a stable operating history. Finally, we calculate and present the AEP loss quantification for selected turbines based on the method employed by \cite{Byrne2020}, \cite{Astolfi2021V52Aging}, and \cite{Astolfi2022}, and the sensitivity analysis we employed in \cite{VUGS_AND_BUCH_TO_THE_TOP}. Our extensive study combines many established methods from the field with recent innovations. It also works to create an expansive end-to-end framework for AEP loss quantification.

\section{Dataset}\label{sec:dataset}
The dataset employed in this study is sourced from Vestas and encompasses SCADA signals from Vestas' wind turbines. In total, the dataset includes four offshore WFs and eleven onshore WFs. Unlike our previous study \cite{VUGS_AND_BUCH_TO_THE_TOP}, the present work does not limit the number of WFs in the analysis prior to data filtering or preprocessing. Each record is a measurement of SCADA signals that capture different WT phenomena, such as the various operation modes, and realize the dependent variable, namely the turbine's energy production, measured in kilowatts (kW). All data originate from turbines operated by Vestas' customers. Vestas performs an unspecified aggregation process to protect proprietary information and ensure data privacy. Certain data attributes are intentionally excluded to maintain confidentiality. For example, no direct locational metadata is provided. The meteorological data are limited to the temperature, wind speed, and wind direction. Consequently, certain visualizations in this work omit axis ticks to preserve confidentiality. The SCADA data comprise 15 distinct attributes, each serving a specific descriptive or operational purpose. The \verb|AmbTemp| attribute represents the ambient temperature at the time of measurement, measured in Celsius. The attributes \verb|BladeLoadA|, \verb|BladeLoadB|, and \verb|BladeLoadC| denote the pressure exerted on each blade due to wind force. \verb|GridPower| indicates the produced power measured in kW, while \verb|WindSpeed| measures the wind velocity in meters per second (m/s). The attributes \verb|PitchAngleA|, \verb|PitchAngleB|, and \verb|PitchAngleC| capture the angular orientation of each blade. Temporal information is recorded in the \verb|Time| attribute, which specifies the moment of measurement and is aggregated in ten-minute intervals. The identifier \verb|TurbineId| uniquely references individual turbines. \verb|WSE| (Wind Speed Estimated) provides a computed wind-speed value inferred from the measured power output (\verb|GridPower|) and is therefore not included further in this study. In addition, \verb|WdAbs| denotes the absolute wind direction, whereas \verb|WindDirRel| measures the relative wind direction.

\begin{figure}[t]
    \centering
    \includegraphics[width=1\linewidth]{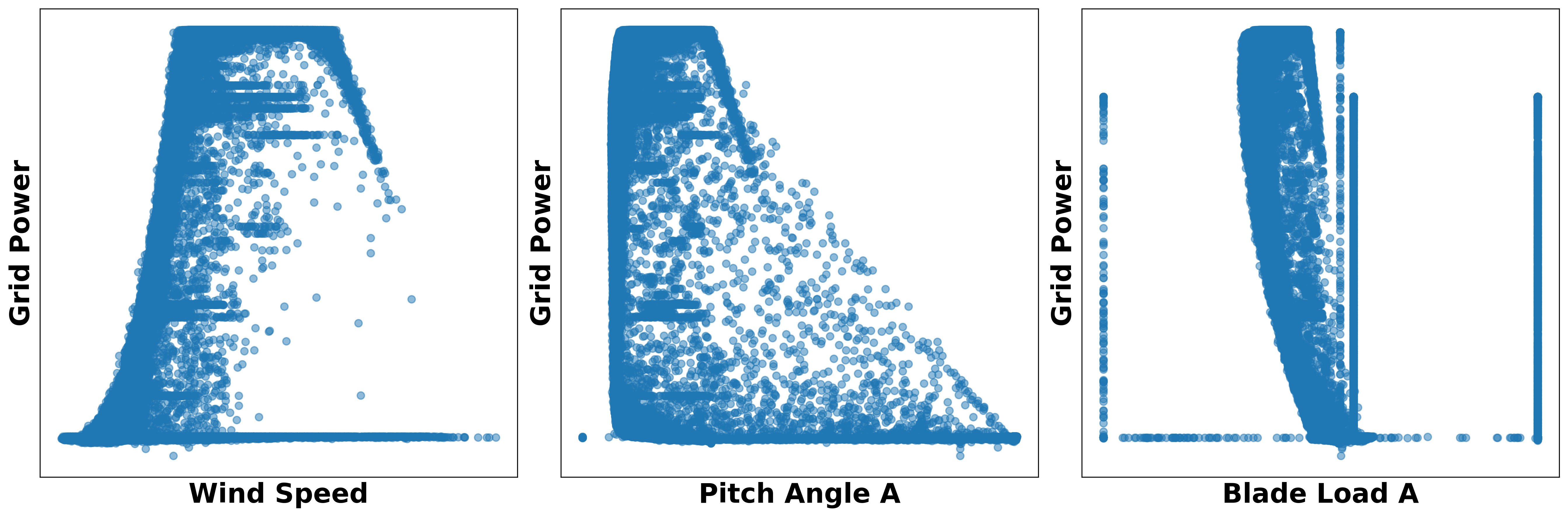}
    \caption{Gridpower relationships for WF~Id~1 Turbine~Id~6}
    \label{fig:p1t6-gridpower-relations}
\end{figure}

\begin{figure}[t]
    \centering
    \includegraphics[width=1\linewidth]{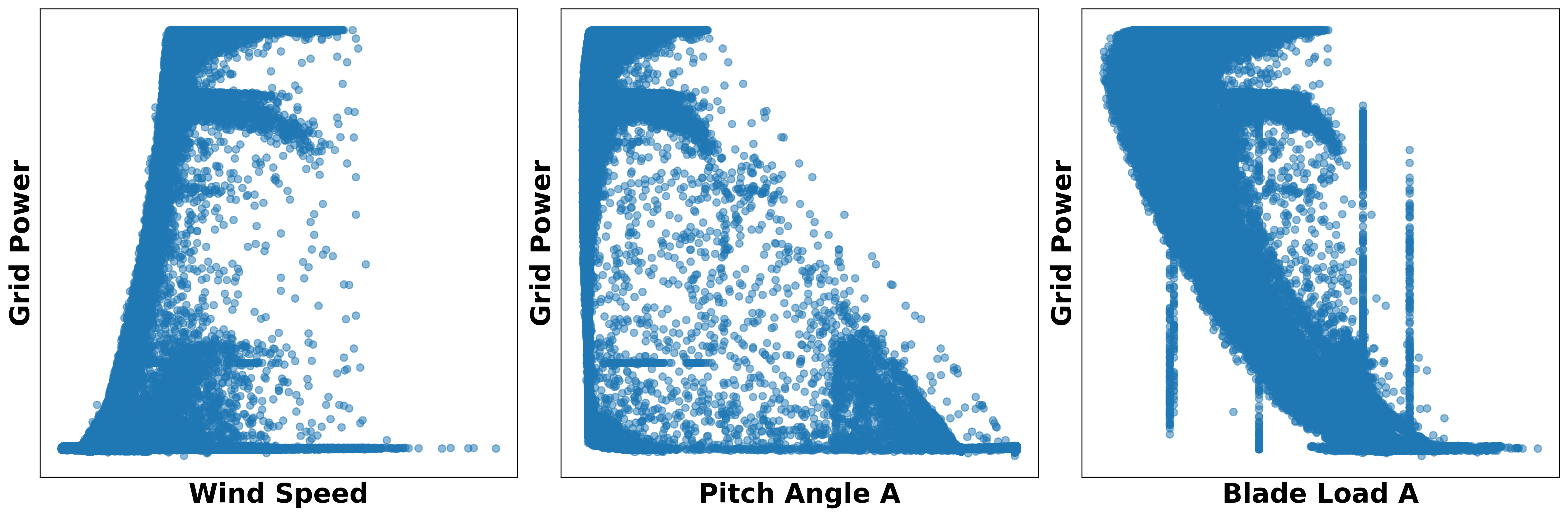}
    \caption{Gridpower relationships for WF~Id~11 Turbine~Id~74}
    \label{fig:p11t74-gridpower-relations}
\end{figure}

\Cref{tab:turbine_data} provides an overview of the WFs supplied by Vestas, including an indicator of offshore or onshore status, the SCADA data time span, the number of turbines in each WF, the total number of data points in the WF, and the average number of data points per turbine. The data we investigate ranges from 2015 for the earliest observations to 2024 for the latest, with varying starting periods subject to WF. The contiguous Id ranges indicate that individual farms host between five and ten turbines. \Cref{fig:p1t6-gridpower-relations} and \Cref{fig:p11t74-gridpower-relations} illustrate the raw data for the offshore WT with Id 6 and onshore WT with Id 74, respectively, and the relationships between \verb|GridPower| and its most correlated variables.


\subsection{Anomalies in SCADA data}
\begin{figure}
    \centering
    \includegraphics[width=1\linewidth]{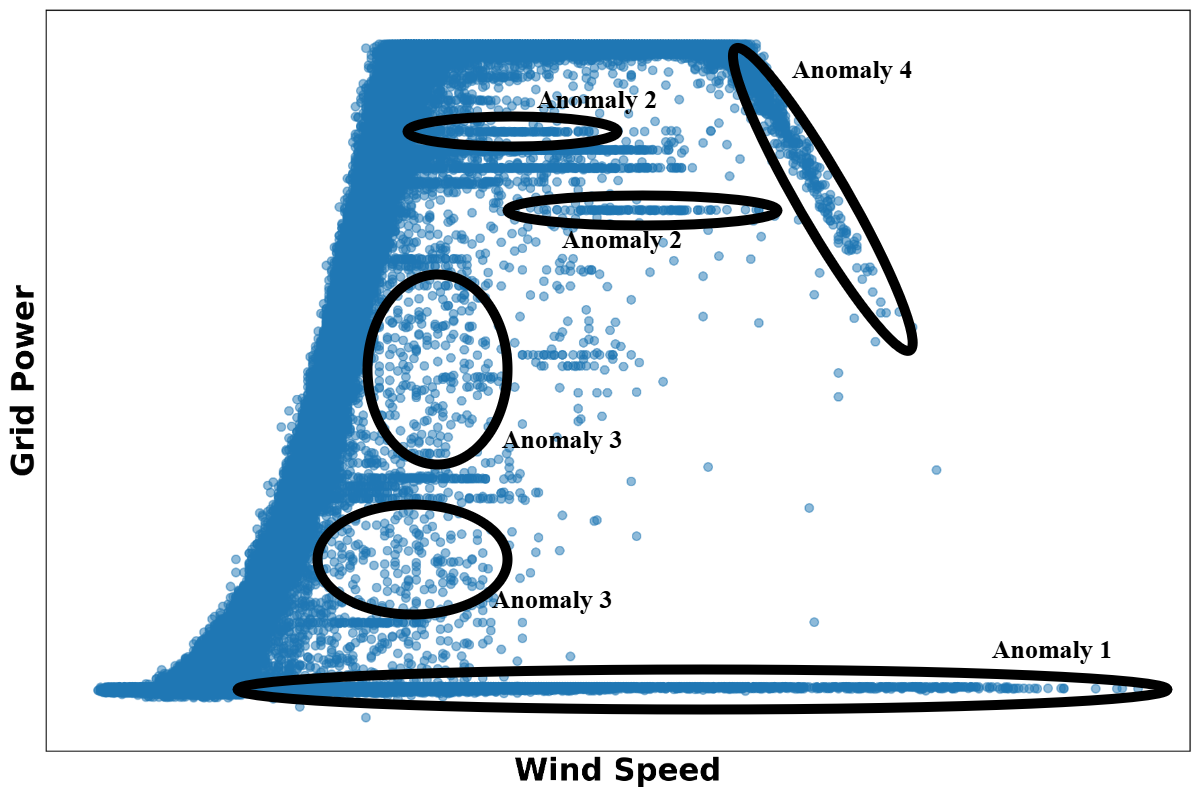}
    \caption{PC Anomaly types - WF~Id~1 Turbine~Id~6}
    \label{fig:p1t6-pc-anomaly}
\end{figure}

The spread in the average records per turbine and the vertical lines seen in the raw data suggest missing data, prolonged outages, or anomalous behavior. Anomalies are defined as instances that do not conform to the patterns present in the dataset \cite{chandola2009anomaly}. These anomalies suggest that the observations in question have been generated by a mechanism fundamentally different from that, producing the notion of NB data. It is, therefore, essential to identify and remove such anomalies to avoid biasing the relationships under investigation. Within WT SCADA datasets, taxonomies typically delineate three principal anomaly classes \cite{morrison2022anomaly}\cite{lin2020wind}. However, empirical evidence reveals a fourth category absent from the referenced literature. Any anomaly detection (AD) method must be designed with these categories in mind, described below and illustrated in \Cref{fig:p1t6-pc-anomaly}.

\begin{itemize}
    \item Anomaly type 1 denotes intervals during which the turbine generates no electrical power even though the wind speed surpasses the cut-in threshold, typically reflecting downtime scheduled by the operator. Within the framework of AD, such events are classified as contextual anomalies \cite{chandola2009anomaly}. Although each feature value appears to be NB when examined in isolation, their collective configuration deviates markedly from the expected operational context.

    \item Anomaly type 2 corresponds to consistently sustaining a positive power production below the turbine's rated capacity, namely Below Rated Power (BRP). This is often a consequence of curtailment imposed by the operator and is within the AD taxonomy classified as contextual anomalies \cite{morrison2022anomaly}.

    \item Anomaly type 3 encompasses observations sporadically distributed across the feature space without any notable structure. As indicated in \cite{lin2020wind}, such irregularities may originate from sensor malfunctions, signal-processing noise, or artifacts introduced by the 10-minute aggregation interval in which the data is provided. The irregularities may also occur during turbine transitions between shutdown and normal operation. These observations are categorized as point anomalies and describe isolated data points whose attributes deviate from the predominant distribution.

    \item Anomaly type 4 represents operating states recorded at high wind speeds near the turbine's rated power region, during which the blades are deliberately pitched out to limit aerodynamic loads. This regime, known as the power derating region, exhibits a monotonic reduction in \verb|GridPower| as blade pitch increases, yielding the characteristic of a downward sloping tail of the power curve. Although such behavior is a protective response or management of load, often invoked when only a subset of turbines is needed to satisfy demand, it deviates from the canonical sigmoidal PC profile. It is, therefore, classified as a contextual anomaly.
\end{itemize}

\Cref{fig:p1t6-pc-anomaly} illustrates anomaly classes commonly observed across the majority of WTs. Detecting such deviations within high-dimensional SCADA data is analytically demanding, yet much of this effort could be prevented through improved data annotation practices. Consequently, we share the statement that:

\begin{quote}
\textit{"Fault, downtime, and curtailment instances should be explicitly labeled in the SCADA by any competent system. It only remains for the user to remove these instances." }\cite{morrison2022anomaly}[p.~475]
\end{quote}
 
Due to the absence of such labeling, as a result of privacy preservation, it is necessary to rigorously define and isolate NB records before implementing CM strategies with ML frameworks that rely on NB training data. \Cref{sec:normality} formalizes our NB definition and the implementation, with \Cref{sec:preliminary_concepts} detailing the methodologies employed to isolate NB observations.
\section{Preliminary Concepts}\label{sec:preliminary_concepts}
This section introduces the reader to the notation used in this paper and provides a background to the methods used subsequently in \Cref{sec:normality}. Consequently, the following outlines and definitions in this section and in \Cref{sec:ml-strategies} will mainly focus on definitions utilized in this paper, giving the reader an understanding of the applied methods. The methods are explained in a vacuum, and their application will be expanded in \Cref{sec:normality}. We assume the reader is familiar with a common ML theoretical framework. Similarly, we assume the reader is familiar with key concepts of probability theory, and we will only cover what is necessary.

\subsection{General Notation}
To keep the theoretical background concise for the following explanations, we introduce the notation used by \cite{bishop2007} and modify references to follow similar conventions. An observation corresponding to a row in the SCADA dataset is denoted as a column vector $\*{x}$, with $D$-dimensional entries. Following that the superscript $\top$ is the transpose, we can for $N$ observations $\*{x}_1,...,\*{x}_N$ with a $D$-dimensional vector $\*x=(\*x_1,\dots,\*x_D)^\top$ construct the data matrix $\*X$ with dimensionality $N \times D$. It follows that the $i^{th}$ element of the $n^{th}$ observation $\*x_n$ is the corresponding $n,i$ element for $\*X$ \cite{bishop2007}. When we explain ML models from a deterministic viewpoint in \Cref{sec:ml-strategies}, the task at hand is learning a model $f_*$'s parameters $\'\theta$ over some data $\*x$ that maps predictions as $\hat{\*y}=f_*(\*x;\'\theta)$ \cite{Goodfellow-et-al-2016}. We mostly use and substitute $_*$ in $f_*$ for indexing or referencing specific models, and we use similar indexing for some of the applied NB methods. When we mention predictions in \Cref{sec:related}, we refer to the outcomes of a model, given an input $\*x$, $f_*(\*x)=\hat{\*y}$. Furthermore, a residual is the difference between an observed target value $\*y$ and its corresponding prediction
\begin{align}
\label{eq:residual}
    \*r &= \*y- f_*(\*x) \\
    &=\*y-\hat{\*y}, \nonumber
\end{align}
and a residual distribution is obtained via a data matrix 
\begin{align}
\label{eq:residual_dist}
    \*R &= \*Y-f_*(\*X) \\
    &=\*Y-\hat{\*Y}\nonumber
\end{align}
In our case, we only predict the \verb|GridPower|, therefore $\*Y$ and $\hat{\*Y}$ are $N\times1$ row vectors. For the probabilistic model, the reader should assume that $p(\*x)$ is the probability density function of the joint probability density function of the random vector $\*X=(X_1,\dots,X_D)^\top$.

\subsection{Box-Plot Rule}
\label{subsec:box_plots}
The box-plot rule, also known as Tukey fences, is often used in this research and is widely applicable because it is non-parametric. The box-plot is a standard that displays a univariate distribution, and not considering outliers, it is described by five key characterizations \cite{wikipedia_box_plot}:
\begin{itemize}
    \item Minimum, $0th$ percentile, $Q_0$.
    \item First quartile, $25th$ percentile, $Q_1$.
    \item Second quartile, $50th$ percentile, $Q_2$.
    \item Third quartile, $75th$ percentile, $Q_3$.
    \item Maximum, $100th$ percentile, $Q_4$.
\end{itemize}
We utilize the quartiles $Q_1$ and $Q_3$ to establish bounds based on the interquartile range (IQR):
\begin{equation}
\label{eq:IQR}
\text{IQR}=Q_3-Q_1
\end{equation}

and the fences are defined by
\begin{align}
\label{eq:tukey_fences}
    F_{upper} &= Q_3 + 1.5\times\text{IQR} \\
    F_{lower} &= Q_1 + 1.5\times\text{IQR}.\nonumber
\end{align}
An observation $\*x$ is then considered an outlier, if $$\*x< F_{lower}\vee\*x>F_{upper}.$$
For some applications, the density is so extreme that $Q_1 = Q_3$ and \Cref{eq:tukey_fences} will consider all data out of bounds. To ensure we keep the data, we relax the bounds for a less strict version
$$\*x\leq F_{lower}\vee\*x\geq F_{upper}.$$

\subsection{Gaussian Mixture Model}
\label{subsec:GMM_theory}
Mixtures of Gaussian springs from a probabilistic framework as an unsupervised machine learning model, where the assumption is that the underlying data can be modeled with $k$ Gaussians \cite{morrison2022anomaly}. In this study, we work with multidimensional data, and therefore, we need to first define the multivariate Gaussian. Formally, for a $D$-dimensional vector $\*{x}$, the multivariate Gaussian distribution is denoted as 
\begin{equation}
\label{eq:muti_var_gaussian}
    \mathcal{N}(\*{x}|\boldsymbol{\mu},\boldsymbol{\Sigma})=
    \dfrac{1}{2\pi^{D/2}}\dfrac{1}{\boldsymbol{|\Sigma|}}\exp{\left(
    -\dfrac{1}{2}
    (\*{x}-\boldsymbol{\mu})^\top
    \boldsymbol{\Sigma}^{-1}
    (\*{x}-\boldsymbol{\mu})
    \right)},
\end{equation}
where $\boldsymbol{\mu}$ is the $D$-dimensional mean vector and $\boldsymbol{\Sigma}$ is the $D\times D$ covariance matrix with $|\boldsymbol{\Sigma}|$ denoting the determinant of $\boldsymbol{\Sigma}$ \cite{bishop2007}.

The GMM is then defined as a linear superposition of \Cref{eq:muti_var_gaussian}, written on the form
\begin{equation*}
\label{eq:gassian_mixture}
p(\*x)=
\sum_{k=1}^{K}
\pi_k
\mathcal{N}
(
\*{x}|
\boldsymbol{\mu}_k,\boldsymbol{\Sigma}_k
)
\end{equation*}
where 
$\mathcal{N}
(
\*{x}|
\boldsymbol{\mu}_k,\boldsymbol{\Sigma}_k
)$ is a component $k$ and $\pi_k$ are mixing coefficients \cite{bishop2007}. For real probabilities, we have that 
\begin{equation*}
\label{eq:coef_sum}
\sum_{k=1}^{K}\pi_k=1,
\end{equation*}
given that $\mathcal{N}
(
\*{x}|
\boldsymbol{\mu}_k,\boldsymbol{\Sigma}_k
) \geq 0
$, and $\pi_k\geq0$ for all $k$. As suggested by \cite{morrison2022anomaly} and mentioned in \Cref{sec:normality}, we include the GMM for AD, applying the box-plot rule directly on the likelihood of the data belonging to one of the mixtures. By constructing the $N\times D$ data matrix $\*X$, we can express the likelihood of the data belonging to the mixtures by calculating the log-likelihood:
\begin{equation}
\label{eq:mixture_log_likelihood}
\ln{p(\*{X}|\boldsymbol{\pi},\boldsymbol{\mu},\boldsymbol{\Sigma})} =
\sum^N_{n=1}
\ln{
\left(
\sum^K_{k=1}
\pi_k
\mathcal{N}(
\*{x}_n|
\boldsymbol{\mu}_k,\boldsymbol{\Sigma}_k
) 
\right)
}
\end{equation}
Practically, this is implemented by \verb|sklearn| \cite{scikit-learn}, and we call the \verb|score_sample| function like \cite{morrison2022anomaly}.

\subsection{Mahalanobis Outlier Detection}
\label{subsec:Mahalanobis_theory}
The Mahalanobis distance is an effective metric for identifying outliers in multivariate space. The main feature is its inclusion of the covariance, considering the correlations between variables. For a point $\*x$, we define the Mahalanobis distance as
\begin{equation*}
\label{eq:mahalanobis_distance}
D^2_{Mahalanobis}(\*{x)} =  
(\*{x}-\boldsymbol{\mu})^\top
\Sigma^{-1}
(\*{x}-\boldsymbol{\mu}),
\end{equation*}
and is the distance from $\*{x}$ to $\boldsymbol{\mu}$ \cite{bishop2007}\cite{holbert_mahalanobis_2021}. What makes the Mahalanobis distance unique is that it follows a chi-squared distribution ($\chi_k^2$) with $k$ degrees of freedom \cite{holbert_mahalanobis_2021}\cite{wikipedia_mahalanobis}.

Similar to \cite{holbert_mahalanobis_2021} and \cite{cansiz2021multivariate}, we test the Mahalanobis distance against the $\chi^2$ distribution with significance level $\alpha$. We set the critical value at $\chi^2_{crit}=\chi^2_{k,1-\alpha}$ with $k$ degrees of freedom, and the decision rule is then calculated as \cite{holbert_mahalanobis_2021}
\begin{equation}
\label{eq:mahalanobis_threshold}
D^2_{Mahalanobis}(\*{x)}>\chi^2_{crit}.
\end{equation}

Any observation $\*x$ with a distance greater than the critical value is flagged as an outlier and removed.

\subsection{Predictive Power Score}
The PPS lays the foundation for much of the data validation in this study. First introduced by \cite{wetschoreck_krabel_krishnamurthy_2020}, the PPS is a metric used to evaluate the predictive strength of one variable on another. The PPS is an alternative to classical statistical tools like Pearson correlation, which mainly identifies linear relationships. Instead, the PPS detects non-linear and asymmetric relationships between variables that linear tools cannot uncover. By considering a simple version of the PPS, we can construct two bivariate data sets $\*X_{train}$ and $\*X_{test}$, with explainable variable $e$ and target variable $t$, and train two models on the training data $f_{model}$ and $f_{naive}$. The naive model $f_{naive}$ is any naive estimate we define, like a random guess or the median, where the $f_{model}$ is typically an ML model with expectations of better predictions than the naive model. With an error metric $M_{\epsilon}$, the normalized error can be expressed as 
\begin{equation}
\label{eq:norm_error}
\mathcal{E}=\dfrac{M_\epsilon(f_{model}(\*X_{test}))}{M_\epsilon(f_{naive}(\*X_{test}))},
\end{equation}
with $M_\epsilon\geq0$ and the PPS for $e$ on $t$ can then be expressed as

\begin{equation}
\label{eq:PPS_max}
    PPS(e,t)= \max(0, 1-\mathcal{E})
\end{equation}

The constraints of \Cref{eq:PPS_max} ensure the PPS score of $e$ on $t$ will always fall between $0$ and $1$, where $1$ signals a strong predictive power. The model choice for \Cref{eq:norm_error} is mainly up to the user. The authors of PPS \cite{wetschoreck_krabel_krishnamurthy_2020} utilize the median as a predictor for the naive model $f_{naive}$ and a decision tree for the model $f_{model}$. This study uses the same models as \cite{wetschoreck_krabel_krishnamurthy_2020}.  
Furthermore, it is derived from a K-fold cross-validation framework to ensure the PPS is reflective and generalized. To properly employ the PPS, we re-implement the library to work within our temporal K-fold cross-validation framework further delineated in \Cref{sec:ml-strategies}, and provides a similar CM application to \cite{bonacina2022use}, that showcase a drop in PPS can be directly linked to system performance reduction up to eight months before a repair is done.

\subsection{k-Selection Strategy} \label{subsec:k-selection-strategy}
A common method for selecting the number of components in GMM is the Bayesian Information Criterion or the curve \textit{elbow}, also employed by \cite{morrison2022anomaly}. However, the goal of applying GMM will be to maximize the PPS. Therefore, we cast the problem of finding the correct number of components as a trade-off between the gain in PPS against the proportion of data flagged as outliers. For $N$ observations, $N_{\delta,k}$ is the proportion of removed data after applying the box-plot rule to the outcome of \Cref{eq:mixture_log_likelihood} for a $k$-component. Calculated as number of observations pre-filtering $N_a$ against number of observations  post-filtering $N_b$, the $N_{\delta,k}$ computed as
$$N_{\delta,k}=1-N_{b,k} / N_{a}$$ 
Generally, for a given period with a set of possible $k$'s $1,\dots,K$ with adjustable weight $\alpha$, the best mixture is found by 

\begin{equation}
\label{eq:kselection}
k_*=\arg\max_{k\in K}(\alpha\cdot\text{PPS}_k-(1-\alpha)\cdot N_{\delta,k}),
\end{equation}
where $0\leq\alpha\leq1$. In this study, we pick $\alpha$ based on empirical evaluation and feedback from Vestas. However, if a clear objective can be identified between the trade-offs of PPS and $N_{\delta,k}$, $\alpha$ can potentially be a learned parameter.

\subsection{Robust Scaler}
When training ML models, it is often a common problem that various features of the dataset are reported in different magnitudes and scales. For example, \cite{deamorim2023scaling} claims that ML algorithms can suffer a reduction in predictive performance due to the algorithms' tendency to rely on dominant features with a high variance that is not necessarily the most informative. To address this issue, scaling procedures like the Min–Max Scaler or the Robust scaler are often employed to reconcile the scales between feature dimensions. Due to the Robust Scaler's ability to mitigate the effects of outliers by centering the data around its median is an obvious choice for this research, where outliers are the main drive \cite{deamorim2023scaling}. 
We can express the Robust Scaler in terms of a $D$-dimensional observation $\*x = (\*x_1,\dots,\*x_D)\top$, with associated $D$-dimensional quantile vectors $\textbf{Q}_1$, $\textbf{Q}_2$ and $\textbf{Q}_3$
\begin{equation}
\label{eq:robust_scaler}
\*x' = \dfrac{\*x-\textbf{Q}_{2}}{\textbf{IQR}},
\end{equation}
where the IQR is calculated as \Cref{eq:IQR} for each respective $D$. We presented unique cases in \Cref{subsec:box_plots}, where some feature dimensions are so dense that $Q_1=Q_3$. When that happens, $\text{IQR}=0$, and we consequently relax the fences on the box-plot rules. However, as seen \Cref{eq:robust_scaler}, $\text{IQR}=0$ will lead to division-by-zero. To tackle this scenario, we utilize \verb|sklearn|'s implementation of the Robust Scaler, which handles division-by-zero, by replacing the IQR with $1$ if $Q_1=Q_3$ \cite{scikit-learn}.
\section{Defining normal behavior}\label{sec:normality}
Studies like \cite{jin2022condition}\cite{meyer2020data}\cite{gill2011wind} define NB as observations that conform to the manufacturer's theoretical PC and remove data points that deviate from this benchmark using visual inspections. Such an approach is imprecise and inefficient \cite{wang2014copula}. Although the theoretical PC exhibits a canonical sigmoidal relationship between wind speed and electrical output, empirical measurements frequently diverge from this PC. Aerodynamic interactions such as wake effects and turbines being in close proximity can affect power production, such that healthy operation may fail to satisfy the NB criterion using the PC. 

\begin{figure}[h]
    \centering
    \includegraphics[width=0.8\linewidth]{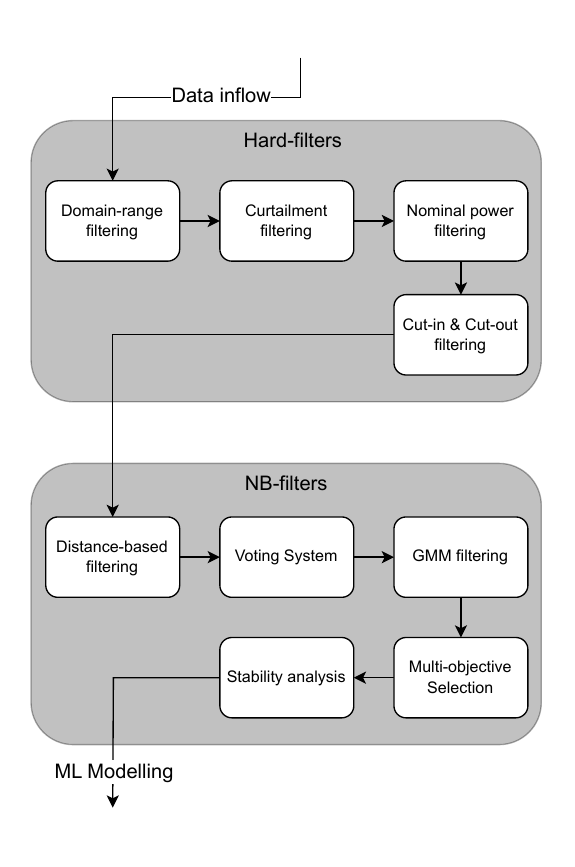}
    \caption{High-level data processing workflow}
    \label{fig:hl-workflow}
\end{figure}

Defining NB can be challenging because each turbine's performance is heavily shaped by its local environment and layout. To address these limitations, the present study adopts a turbine-specific approach to defining normality, grounded in patterns extracted from operational data. 
Instead of relying on a theoretical model, this methodology emphasizes the detection of deviations from each turbine's expected performance, enabling the identification of turbine-specific behavior. The flowchart presented in \Cref{fig:hl-workflow} outlines the general workflow for isolating NB data points and provides a high-level overview of the data preprocessing pipeline. Our pipeline systematically processes the raw SCADA data by applying a sequence of filtering and transformation steps to ensure the identification of operationally representative NB instances.

\subsection{Hard-filters}
A key component of the preprocessing pipeline is data filtering, particularly removing obvious anomalies that could affect the performance of density and distance-based AD methods. In collaboration with a domain expert at Vestas, we developed and implemented a set of domain-range filters designed to exclude outliers and operationally anomalous instances from the dataset. In addition, we also incorporate operational constraints such as deliberately excluding data points corresponding to nominal power output, as such instances typically indicate optimal turbine performance with no associated energy production loss. The exclusion of nominal power data constitutes BRP, as described in \Cref{sec:dataset}. Operating at nominal power means that the turbine functions at its rated capacity given the environmental conditions, effectively representing a sustainable upper limit of energy output. Therefore, we define hard-filters as comprising domain-range filters and operational constraints essential for delineating NB's boundaries. Specifically, these hard-filters refer to the steps in the upper box of \Cref{fig:hl-workflow} and are detailed in \Cref{tbl:hard-filter}. 

\begin{table}[t]
\renewcommand{\arraystretch}{1.3}
\centering
\caption{Summary of hard-filters.}
\label{tbl:hard-filter}
\begin{tabular}{P{2.5cm}|P{5.5cm}}
{Filter Name}           & {Filter Definition} \\
\hline
Domain-range filters    & AmbTemp $\geq$ -10, \\
                        & BladeLoadA, BladeLoadB, BladeLoadC~$\leq$~0, \\
                        & Year $<$ 2024 \\
Curtailment             & PitchAngleA, PitchAngleB, and PitchAngleC~$\leq$~0  \\
Nominal Power           & GridPower $\leq$ 95\% of Max GridPower \\
Cut-in                  & (WindSpeed $\geq$ 5) OR (GridPower $\geq$ 5\% of Max GridPower) \\
Cut-out                 & WindSpeed $\leq$ 20       \\
\end{tabular}
\end{table}

\begin{itemize}
    \item Domain-range filters define the rules used to remove obvious outliers from the dataset. In collaboration with Vestas, we determined that ambient temperatures below –10 degrees Celsius are regarded as rare occurrences for the given WFs and should, therefore, be treated as erroneous data. For blade load measurements, which capture the bending forces exerted by the wind, we expect negative values due to the backward bending of the blades under normal operating conditions. Consequently, we classify any positive blade load values as anomalous. Finally, we exclude data from the year 2024 due to an insufficient number of observations.

    \item Curtailment are periods where electricity output from wind turbines is deliberately reduced, despite their capability to generate more power under existing wind conditions. Thus, this filter describes an operational limitation. Additionally, the curtailment includes observations at or beyond nominal power, as these state that the turbine is operating at its rated maximum capacity, effectively indicating a sustained energy output.

    \item Nominal power refers to the region of the sigmoidal PC where the turbine operates at its maximum rated capacity. While the curtailment filter removes a subset of such data points, it does not capture all instances of nominal power operation. To address this limitation, we implement an additional operational filter specifically designed to identify and exclude remaining nominal power data points.

    \item Cut-in and cut-out define the lower and upper operational wind speed thresholds for a WT. The cut-in speed specifies the minimum wind speed at which the turbine begins generating electrical power. In contrast, the cut-out speed marks the maximum allowable wind speed beyond which the turbine shuts down to prevent mechanical damage from excessive aerodynamic loads.
\end{itemize}

\begin{figure}
    \centering
    \includegraphics[width=0.75\linewidth]{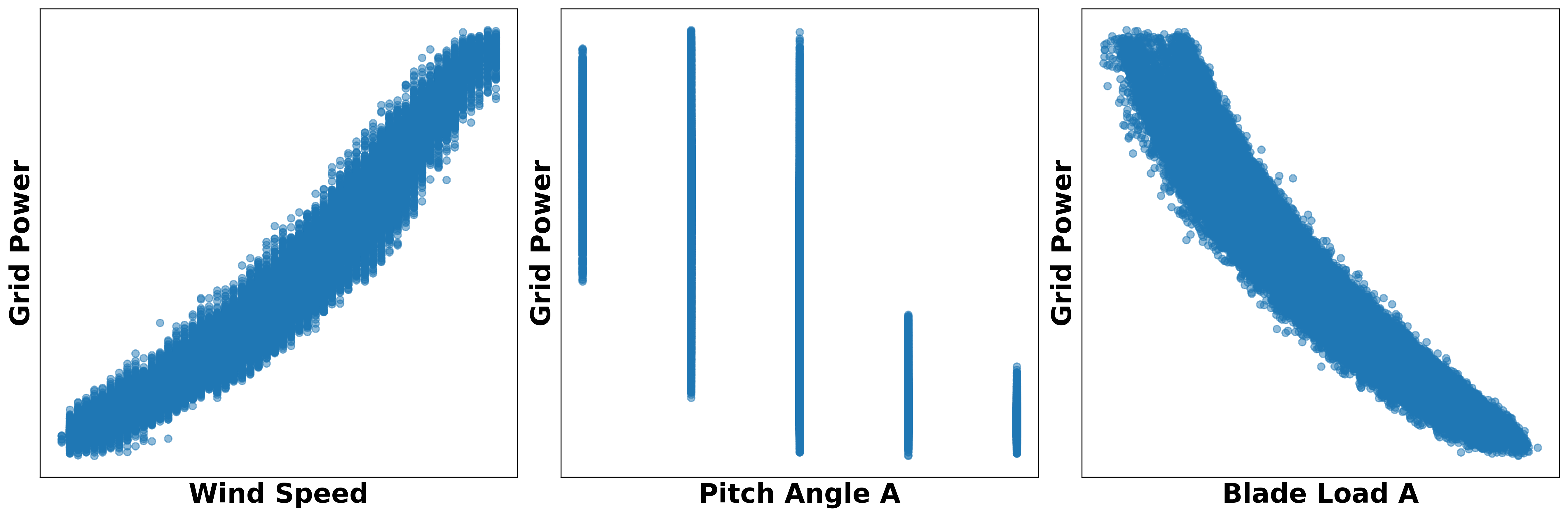}
    \caption{Gridpower relationships for WF Id 1 Turbine Id 6 after filtering}
    \label{fig:p1t6-nb}
\end{figure}
\begin{figure}
    \centering
    \includegraphics[width=0.75\linewidth]{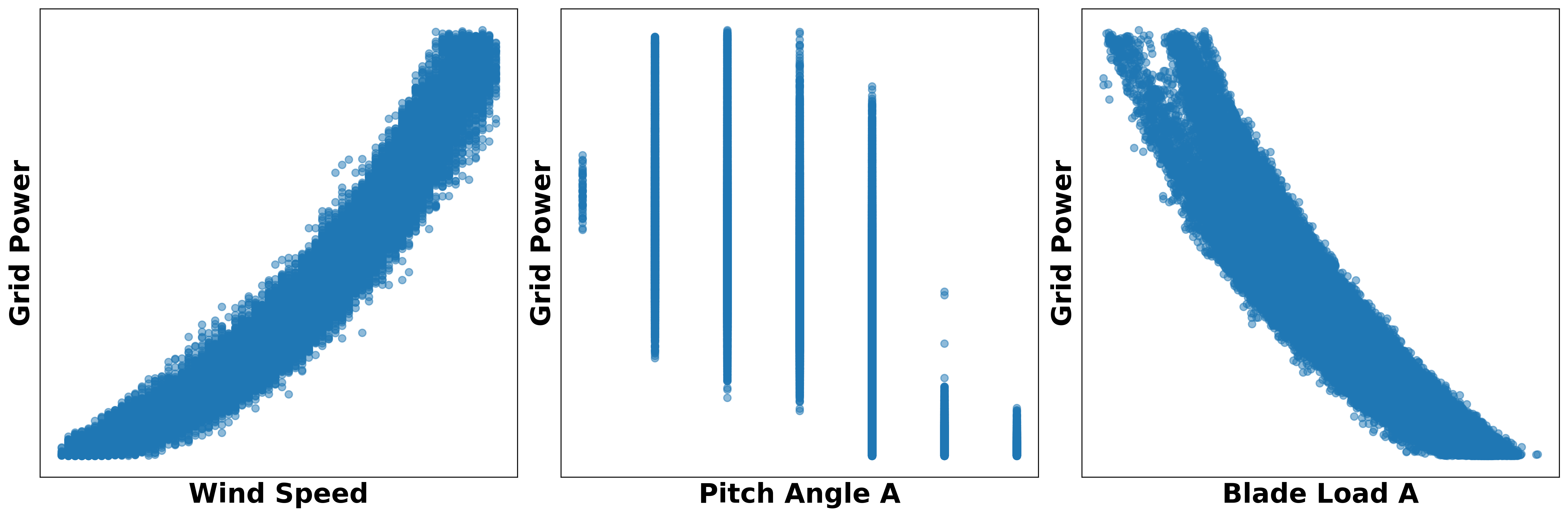}
    \caption{Gridpower relationships for WF Id 11 Turbine Id 74 after filtering}
    \label{fig:p11t74-nb}
\end{figure}

\subsection{NB-filters}
While the hard-filters effectively remove a significant subset of anomalies, such as the extreme values evident in the vertical patterns observed in \Cref{fig:p1t6-gridpower-relations} and \Cref{fig:p11t74-gridpower-relations}, they are insufficient for comprehensive anomaly detection, as demonstrated in our prior work \cite{VUGS_AND_BUCH_TO_THE_TOP}. Similarly, applying a single density-based AD method, such as Local Outlier Factor, was also inadequate in capturing the full spectrum of anomalous behavior \cite{VUGS_AND_BUCH_TO_THE_TOP}. To address these limitations, we introduce NB-filters, which comprise an ensemble of AD methods, including density- and distance-based methods and a PPS-based AD framework. These methods increase the number of anomalous data points identified and removed, resulting in a more accurate and turbine-specific representation of NB. This task falls within the domain of unsupervised learning, where ground-truth labels are not available to assess the accuracy of the AD methods directly. Therefore, validation relies on a combination of strategies, including visual inspection and empirical analysis, expert confirmation from Vesta's domain specialist, improvements in downstream machine learning model performance, and assessment of residual distributions for conformity to a normal distribution. The effects of applying the hard-filters and NB-filters are illustrated in \Cref{fig:p1t6-nb} and \Cref{fig:p11t74-nb} showcasing a much smoother PC similar to the sigmoidal shape in the theoretical PC. A detailed evaluation of the filtration results is provided in \Cref{sec:results}.

The NB-filters introduced in this study constitute one of the primary extensions of our previous work to further refine the isolation of NB data from SCADA signals. The NB-filters refer to the lower box in \Cref{fig:hl-workflow} and is a framework composed of an ensemble of AD methods designed to improve the identification of NB operational data in a scalable and generalizable way, applicable across WTs and WFs. We achieve this by applying NB-filters on each turbine, allowing NB to be defined relative to its operational characteristics. This turbine-specific approach gives a more accurate understanding of what constitutes normal operation. Additionally, the AD methods employed in this study are more computationally efficient compared to algorithms such as the Local Outlier Factor. The AD methods require less computational resources, making them well-suited for large-scale SCADA data. Finally, this method maintains privacy, as it doesn't rely on geolocation data for WTs. This omission restricts the integration of weather parameters, such as precipitation rates or other environmental factors, commonly leveraged in prior research. Studies like those by \cite{visbech2023introducing} and \cite{panthi2023quantification} have effectively utilized weather information to connect environmental conditions with erosion severity and energy losses. By incorporating site-specific meteorological information, these approaches also accurately enhance the ability to isolate NB within turbine operational data. This study exclusively utilizes SCADA signals for inferring performance degradation, making it an easily scalable and cost-effective approach to CM, particularly in scenarios where external data sources are unavailable or impractical to obtain.

The NB-filtering framework is constructed on five sequential steps, each building on the previous stage to refine the dataset progressively. As the sequence advances, the AD methods employed are increasingly sensitive to data quality and more dependent on the prior removal of preliminary outliers, especially density-based methods. Consequently, the initial AD steps are robust to noise, but the later steps assume a cleaner dataset for anomaly detection. The effectiveness of the NB-filters depends on the application of the preceding hard-filters. The complete sequence of the five NB-filtering steps is illustrated in \Cref{fig:hl-workflow} and follows the order presented in the figure.

\medskip

\subsubsection{Distance-based filtering} 
The distance-based filtering methods employed in this study are designed to detect and eliminate statistical outliers from the dataset. 
\begin{enumerate}[label=\roman*.]
    \item The first AD method applied is the Interquartile Range (IQR) method, a non-parametric approach grounded in descriptive statistics. The IQR-based filtering is applied specifically to the pitch angle measurements to reduce the risk of bias for subsequent AD methods. Importantly, the IQR method is particularly well-suited to empirical research and industrial applications because it does not rely on assumptions about the underlying data distribution.

    \item In the second stage of the filtering process, we employ a multivariate outlier detection technique based on the Mahalanobis distance within a bivariate normal framework. This method identifies and removes anomalous relationships between pitch angle measurements and wind speed, two variables expected to exhibit correlation under normal operating conditions. For each pitch angle variable, the joint distribution with wind speed is modeled using the empirical mean vector and covariance matrix. The Mahalanobis distance is computed for each observation relative to this bivariate distribution. Unlike the Euclidean distance, the Mahalanobis distance accounts for the data's scale, orientation, and correlation, making it particularly well-suited for detecting multivariate outliers in correlated SCADA data. 
    
    To formally assess outlier status, we apply a statistical hypothesis test based on the chi-squared distribution with two degrees of freedom, corresponding to the two-dimensional feature space. The significance level from \Cref{subsec:Mahalanobis_theory} determines the filtering threshold and observations with a distance greater than this threshold are filtered as outliers. We apply the test across all pitch angle dimensions. An observation is retained if it's statistically consistent with the threshold for all pitch angles and wind speed pairings. This criterion ensures higher confidence in the retained data, mitigating the influence of sensor noise, measurement drift, and operational anomalies.

    \item Thirdly, a hierarchical IQR filtering method is applied to wind speed measurements, conditioned on discretized pitch angle values. This approach is designed to account for the contextual variability in wind speed distributions that originate from differing pitch angle configurations, which often correspond to distinct operational states or environmental conditions. The SCADA data is split into hierarchical groups, each representing a subset of observations characterized by a specific pitch angle configuration. The Q1 and Q3 of wind speed are computed within each group, and the corresponding IQR is derived. Tukey fences are then applied to identify and remove wind speed outliers that fall significantly outside the central range of the subgroup's distribution. The method respects the underlying heterogeneity in turbine behavior by applying IQR filtering within pitch-specific subgroups rather than globally. This context-aware approach is particularly effective in identifying anomalies associated with operational derating, a condition that previous distance-based AD methods failed to detect reliably. At the same time, the hierarchical filtering preserves the internal variability that characterizes normal turbine operation.
\end{enumerate}

Removing anomalous values from the pitch angles reduces the risk of bias in subsequent AD methods and helps further refine what constitutes NB.

\medskip

\subsubsection{Voting System}
Even though there is a high correlation between the pitch angle measurements (\verb|PitchAngleA|, \verb|PitchAngleB|, and \verb|PitchAngleC|) and the blade load measurements (\verb|BladeLoadA|, \verb|BladeLoadB|, and \verb|BladeLoadC|), that there were numerous cases where sensor readings diverged significantly. As a result, we developed a consensus-based voting system that leverages the PPS. Importantly, the PPS framework focuses on the predictive relationship with the target rather than absolute agreement among sensor values, thus enabling a more robust assessment of sensor reliability. The proposed voting system implements a statistical anomaly detection framework to identify and manage systematic issues in multi-sensor measurement configurations, such as sensor degradation and inter-sensor inconsistencies. These are common challenges in CM systems operating in real-world environments. The methodology proceeds through a structured four-step process that combines distance-based statistical evaluation with consensus-based decision logic.

\begin{enumerate}[label=\roman*.]
    \item The system begins by ingesting SCADA data that has already undergone the prior anomaly detection and filtering from previous steps, thereby establishing a baseline dataset.

    \item A voting mechanism is instantiated with a distance threshold for anomaly detection sensitivity, a removal strategy for handling identified anomalies, and period-specific modification capabilities. The threshold defines the maximum permissible statistical distance between sensors before triggering anomaly classification. The voting mechanism can operate in either strict or non-strict mode. In strict mode, any disagreement among sensors triggers the time interval to be flagged as anomalous. In non-strict mode, anomalies are flagged only when a majority of sensors disagree or when one sensor receives votes from at least two other sensors within the same time interval. This study applies the strict mode as a rigorous removal strategy is needed to capture NB properly.

    \item The algorithm iteratively processes each turbine's multi-dimensional sensor data, comparing homogeneous sensor groups, specifically pitch angle and blade load sensors. Using distance-based metrics, it quantifies inter-sensor deviations. Under normal operational conditions, sensors within the same group are expected to produce concordant measurements, and significant deviations suggest sensor degradation or errors. When the calculated distance exceeds the threshold, a vote is registered against the outlying sensor.
\end{enumerate}

The voting system addresses the practical engineering challenge of maintaining data quality in complex multi-sensor systems where individual sensor failures or calibration drift can compromise entire datasets. This voting-based approach provides a statistically principled method for distinguishing between legitimate operational variations and systematic measurement errors, essential for accurate CM applications.

\medskip

\subsubsection{GMM filtering}
We employ a Gaussian Mixture Model (GMM) approach, as proposed in \cite{morrison2022anomaly}, to probabilistically model anomalous behavior in WT SCADA data. The GMM framework treats the feature space as a composition of $k$ Gaussian components, each representing a distinct normal operation mode. Observations associated with high-density regions of these components are considered representative of NB, while those located in low-probability regions are considered anomalous. Consequently, the GMM is applied specifically to the sigmoidal ramp-up region, where anomaly detection is especially relevant to determining energy loss.

In contrast to distance-based AD methods, which often struggle with stacked high-density data regions, the probabilistic approach of GMM makes it more resilient to stacked data \cite{morrison2022anomaly}. One of the key advantages of the GMM lies in its unsupervised yet interpretable structure. The likelihood of each data point under the fitted mixture model can be explicitly computed, enabling a transparent and data-driven decision rule. This study uses an IQR threshold applied to likelihood values to identify and remove anomalous observations systematically. The GMM-based anomaly detection framework proceeds as follows.

\begin{enumerate}[label=\roman*.]
    \item First, feature selection is performed to retain only variables that exhibit strong statistical relationships with the target variable \verb|GridPower|. Specifically, the retained features include \verb|BladeLoadA|, \verb|BladeLoadB|, \verb|BladeLoadC|, and \verb|WindSpeed|. Variables such as pitch angles are excluded at this stage, as their explanatory power diminishes after the previous filtering steps, wherein a small number of discretized pitch angle values span the entire power output space, as illustrated in \Cref{fig:p1t6-nb} and \Cref{fig:p11t74-nb}. This dimensionality reduction step improves the GMM's capacity to identify meaningful operational patterns by eliminating redundant or low-informative variables that may introduce noise or spurious correlations. Before applying the GMM, we scale the data with the Robust Scaler defined by \Cref{eq:robust_scaler}.

    \item At the core of the methodology is an iterative GMM fitting procedure, where models with $k = 1$ to $k = 5$ mixture components are evaluated. When visualizing the PPS for various $k$, as in \Cref{fig:avg_pps_NB_vs_HF}, the $k = 0$ case serves as a baseline, representing the absence of mixture-based anomaly filtering. For $k \geq 1$, each model fits $k$ Gaussian components to the data, thereby capturing natural clustering structures within the high-dimensional operational space. We calculate the log-likelihood by \Cref{eq:mixture_log_likelihood} for observations belonging to a component and filter using the box-plot rule similar to \cite{morrison2022anomaly}. The PPS is computed for each GMM configuration.
\end{enumerate}

\medskip

\subsubsection{Multi-objective Selection}
A commonly used approach for selecting the number of Gaussian components in mixture models is the Bayesian Information Criterion \cite{morrison2022anomaly}. However, in the context of this study, where data are drawn from numerous WFs and WTs, each exhibiting distinct definitions of NB, a more adaptive and context-aware method is required. Therefore, we propose a dynamic selection strategy incorporating the PPS and a data retention measure. The aim is to select a value of $k$ that simultaneously enhances the PPS, reflecting improved relationships between input variables and the target variable \verb|GridPower| while minimizing the proportion of data removed during filtering. This results in a multi-objective selection criterion that balances data quality with quantity. The dynamic $k$-selection methodology is designed to account for the seasonal and operational variability inherent in SCADA data by optimizing the number of mixture components within discrete quarterly time periods. The principal innovation lies in the implementation of a multi-objective optimization framework for determining the optimal number of Gaussian components, $k$, for each time period. Rather than applying a static, global value of $k$, the method adaptively selects $k$ based on the local characteristics of the data in each temporal window. This optimization employs a weighted indicator to evaluate the trade-off between two competing objectives.

\begin{enumerate}[label=\roman*.]
    \item Maximizing the PPS, which serves as a proxy for the predictive quality of the retained data. 

    \item Minimizing the proportion of data removed, denoted $N_{\delta}$, to preserve as much information as possible. 
\end{enumerate}

The weighting scheme enables tuning of the relative importance of these objectives according to analytical or operational priorities. By dynamically adjusting the number of mixture components, the proposed method achieves an optimal balance between anomaly detection and data availability, tailored to the evolving characteristics of turbine operation over time. An example of identifying the optimal $k$, can be seen in \Cref{appendix:k-selection}

\medskip

\subsubsection{Stability Analysis} \label{subsec:stability_analysis}
The stability-based selection methodology implements a temporally grounded performance assessment to identify WTs that exhibit consistent operational behavior suitable for longitudinal comparative analysis. This approach addresses the challenge of distinguishing between turbines with stable baseline characteristics and those exhibiting significant temporal variability, which may introduce confounding effects into downstream analyses. The principle of the selection process is temporal consistency evaluation, operationalized through a multi-criteria assessment of turbine performance over time.

We define the stable operational period at a quarterly time interval that meets specific threshold values based on the PPS. A stable year is defined as a calendar year where all four quarters meet these stability criteria. The criteria for determining stability are as follows: \begin{enumerate}[label=\roman*.] 

    \item The methodology begins with applying a threshold criterion based on the PPS, establishing the minimum required level of predictive power between explanatory variables and the target variable, \verb|GridPower|. The criterion guarantees that only periods of a sufficiently strong and stable operational signal are considered for downstream tasks.

    \item To further assess the stability of the WTs operational history, a rolling standard deviation of the PPS is computed across consecutive quarterly periods. A predefined threshold is applied to retain only periods with low predictive variability. This constraint encourages selected intervals to be characterized by consistency rather than temporary fluctuation.
\end{enumerate}

An example of identifying stable periods can be seen in \Cref{appendix:stable-period}. Lastly, we constrain the framework to require temporal continuity. Specifically, stable periods must span at least three consecutive years and not initiate later than 2020. The temporal constraint guarantees enough historical data for model training while preserving future time periods for validation and CM. The resulting data set, comprising only those turbines and time periods meeting the defined criteria, is utilized for subsequent analyses and downstream tasks. The stability method enables drift detection and sensitivity experiments outlined in \Cref{sec:experiments}, where performance deviations can be attributed to operational or environmental changes rather than instability in the reference data.
\section{Machine Learning Strategies}\label{sec:ml-strategies}
The following section will focus on giving the reader an understanding of the applied model architectures in the AEP quantification. Similarly to \Cref{sec:preliminary_concepts}, we assume the reader is familiar with a common ML framework. Therefore, this section provides the architectural details that distinguish the models.

\subsection{XGBoost}\label{subsec:XGB_theory}
Since this investigation builds directly upon our earlier work, we include the explanation for XGBoost previously outlined in \cite[Section V.A]{VUGS_AND_BUCH_TO_THE_TOP}.
Functioning as an ensemble learning technique, XGBoost integrates a sequence of weak learners into a single predictive model. The weak learners utilized by XGBoost are decision trees. Through an iterative process, each successive tree refines its predictions based on the residual errors of its predecessors, thereby improving overall accuracy. This approach enables the model to detect and exploit complex, nonlinear patterns within the data.

At the core of XGBoost lies the optimization of an objective function that balances predictive accuracy and model complexity. The general form of the objective function is given as
\begin{equation*}
    \label{eq:xgb-obj}
    \text{Obj} = \sum_{n=1}^N l(\*y_n, \hat{\*y}_n) + \sum_{k=1}^K \Omega(f_k),
\end{equation*}
where $l(\*y_n, \hat{\*y}_n)$  is the loss function that quantifies the difference between the predicted value $\hat{\*y}_n$ and the actual target $\*y_n$, $f_k$ is a tree, and $\Omega(\cdot)$ is the regularization term that controls the complexity of the trees. The term $K$ is the total number of trees added to the model. Regularization is critical as it penalizes overly complex trees, reducing the risk of overfitting and ensuring the model generalizes well to unseen data. Trees in XGBoost are added in an additive manner, such that the prediction at each step $k$ is updated iteratively as
\begin{equation*}
    \label{eq:xgb-additive}
    \hat{\*y}^{(k)} = \hat{\*y}^{(k-1)} + f_k(\*x),
\end{equation*}
where $f_k$ is the newly added decision tree that fits the negative gradient of the loss function. The final prediction returned by the ensemble is then the sum of tree predictions 
\begin{equation*}
    \hat{\*y}=\sum^K_{k=1}f_k(\*x).
\end{equation*}
The negative gradient guides the model to correct the residuals, ensuring that the model converges efficiently toward the optimal solution. XGBoost computes both the gradient (first-order derivative) and the Hessian (second-order derivative) by second-order Taylor expansion of the loss function at each iteration to improve optimization speed and accuracy. This is reflected by updating the objective function as 
\begin{equation*}
    \text{Obj}^{(k)} = \sum_{n=1}^N \left[l(\*y_n, \hat{\*y}_n^{k-1}) + g_n f_k(\*x_n) + \frac{1}{2} h_n f_k^2(\*x_n) \right] + \Omega(f_k)
\end{equation*}
where $g_n$ is the gradient and $h_n$ is the Hessian of the loss function. This second-order approximation enables XGBoost to handle more general loss functions.

XGBoost's scalability, generalizability, and computational speed make it particularly well suited for NBM of SCADA data. These high-frequency operational measurements often present nonlinear interactions influenced by environmental fluctuations and dynamic operating conditions that XGBoost's flexible and complex composition can capture \cite{singh2022scada}. 

\subsection{Random Forest}
\label{subsec:RF_theory}
Random Forest, introduced by \cite{breiman2001random}, falls under the ensemble category with XGBoost. However, Random Forest distinguishes itself from boosting algorithms on a fundamental level. Where boosting learns from its predecessor's mistakes, Random Forest builds a committee of de-correlated trees and predicts by averaging the committees' response \cite{Hastie2009}. Random Forest achieves this by randomly selecting candidate variables for every available split, directly injecting randomness into the construction of every tree it creates. According to \cite{Hastie2009}, Random Forest is a simple model with similar performance to boosting on many problems. However, as noted by \verb|sklearn| \cite{scikit-learn} and \cite{geurts2006extremely}, the added randomness may not always help with regression problems and, in some cases, worsen its performance. Therefore, by default, \verb|sklearn| does not recommend sampling variables when training trees but instead provides the entire variable selection at every split. Consequently, it should be emphasized for this study that we utilize the \verb|RandomForestRegressor| from \verb|sklearn|; however, by not sampling from variables during the training process, the Random Forest is reduced to bootstrap aggregation (bagging) \cite{Hastie2009}.

Suppose we can obtain some prediction $\*y$ from a model $f_*(\*x)$, trained on training data $\*X$ from some data $\*x_1,\dots,\*x_N$. This gives us a single model $f_*$ \cite{Hastie2009}. Instead, we can opt for a more nuanced view by sampling $N$ observations with replacement, $B$ times from the original $\*X$, providing us with $B$ new variants of training sets $\*X_1^*,\dots\*X_B^*$. By doing so, we bootstrap new distributions, each with its unique representation of the original data, and so will a model trained on each sample. A bagging average is the averaged prediction over the collection of models trained on the bootstrap samples \cite{Hastie2009}. This method reduces the variance of the original model by replacing it with many similar representations of the same model. Specifically, we say for every bootstrap sample $\*X_1^*,\dots,\*X^*_B$, we fit a model $f_b$, and construct our ensemble $f_1,\dots,f_B$. For an input $\*x$, the bagging prediction is then the average across the ensemble
$$\hat{\*y}=\dfrac{1}{B}\sum_{b=1}^B f_b(\*x)$$

To construct bagging trees, we use decision trees as the model choice for training the ensemble. Bagging with trees is interesting because the randomness of the bootstrapping will provide different trees, reducing its generally known weakness of high variance.

\subsection{Multi-layer Perceptron}
\label{subsec:MLP_theory}
The Multi-layer Perceptron (MLP) is a simple case of the broader term deep neural network that implements the feedforward neural network. The MLP has no definite definition, but it is generally used for neural networks that consist of only a few layers. The learning objective of the MLP is approximating some function $f_*$, in our case a regression function that maps the prediction $\hat{\*y}=f_*(\*x;\'\theta)$, by learning the parameters $\'\theta$ \cite{Goodfellow-et-al-2016}. The feedforward network is a logically intuitive name for the process of information flowing through the layers of the network, tied together as different functions from the input layer $\*x$ to the output layer $\hat{\*y}$, with the hidden layers in-between. Generally, we can represent these layers as a chain of functions on one another. Suppose we have three layers, $f_1, f_2$, and $f_3$ for a neural network. We represent its feedforward capabilities by chaining together the functions on the form $f(\*x)=f_3(f_2(f_1(\*x)))$, with $f_1$ being the input layer and $f_3$ being the output layer \cite{Goodfellow-et-al-2016}. 

Our MLP will follow the architecture of \cite{mathew2022estimation} with two hidden layers. Each layer in our network computes a weighted sum of its inputs, adds a bias term, and applies a nonlinear activation function to the result. For a layer $f$ and subsequent layer $f+1$ with input $\*x$, we express the activation as
\begin{align*}
\*h^{(f)}&=g(\*W^{(f)\top}\*x+\*b^{(f)})\\
\*h^{(f+1)}&=g(\*W^{(f+1)\top}\*h^{(f)}+\*b^{(f+1)}),
\end{align*}
 where $\*W$ is the weights of a linear transformation, $\*b$ is the biases and $g$ is a nonlinear activation function \cite{Goodfellow-et-al-2016}. As \cite{mathew2022estimation}, we use Rectified Linear Unit (ReLU) activation functions in the hidden layers, defined as
$$\text{ReLU}(\*x)=\text{max}(0,\*x).$$

The activation function for the output layer is usually picked for a specific task. For regression, we apply the identity function, and we express the estimate as a linear combination of the previous layer $f-1$
$$\hat{\*y}=\*W^{(f)\top}\*h^{(f-1)}+\*b^{(f)},$$
and the entire network is presented as the function of the learned parameters $$\hat{\*y}=f_*(\*x;\'\theta) = \*W^{(F)\top}\*h^{(F-1)}+\*b^{(F)}.$$

\subsection{K-Nearest Neighbors}
\label{subsec:KNN_theory}
K-Nearest Neighbors (KNN) for regression is shown by \cite{Janssens2016} to have strong performance on PC modeling. KNN separates itself from earlier introduced models by not being a learned process. Instead, a KNN model retains all the training data, and a prediction is a direct query upon the retained data, making it well suited for NB modeling. Practically, when we ``fit''  a KNN model, we construct indexing for faster querying, where we can look up a point to its nearest neighbors faster, for example, with a KD-tree \cite{scikit-learn}. 

For finding the nearest neighbors to a query point $\*x_q$, we need some distance measure. We employ Minkowski distance, which can be defined in D-dimensional space for a query point to a point in the original data $\*x$ 
\begin{equation*}
D_{Minkowski}(\*{x}_q, \*{x}) =
\left(
\sum_{d=1}^D
|\*{x}_{q,d}-\*{x}_{d}|^p
\right)^{1/p}
\end{equation*}
with $p=1$ being the Manhattan distance and $p=2$ being the Euclidean distance. For the query $\*{x}_q$, we compute prediction from the set of nearest neighbors $N_K(\*{x}_q)$ and our prediction is then averaged valued of the neighborhood \cite{Hastie2009}
\begin{equation*}
   \hat{\*y}=\dfrac{1}{K}\sum_{i\in N_K(\*{x}_q)}\*y_i
\end{equation*}

\subsection{Metrics}
In this section, we introduce the metrics utilized to measure the correctness of our models, presented in \Cref{sec:results}. A common measurement of errors is the Mean Absolute Error (MAE), which quantifies the average distance of errors between predicted and observed values. Using \Cref{eq:residual}, for $N$ observations, we compute the MAE as
$$\text{MAE}=\dfrac{1}{N}\sum_{n=1}^N|\*r_n|.$$

Similarly, we can compute the Mean Absolute Percentage Error (MAPE)
$$\text{MAPE} = \frac{100}{N} \sum_{n=1}^N \frac{|\*r_n|}{\*y_n}.$$

The MAPE is utilized to reflect relative model errors. Because the power output varies by magnitudes between on and offshore WTs, the MAPE makes it possible to compare turbines across parks within the study. Furthermore, the MAPE can be used as a comparative metric for similar studies where anonymization hinders the direct reporting of power and similar values.

\subsection{Hyperparameter optimization}
\label{subsec:hyper_teori}
For this research, we employ the same hyperparameter optimization strategies as our earlier work \cite{VUGS_AND_BUCH_TO_THE_TOP}, and we, therefore, include the explanation from \cite{VUGS_AND_BUCH_TO_THE_TOP} for the temporal framework applied. K-fold cross-validation is integrated into the training process to guarantee the model's robustness and generalizability. This practice ensures that performance metrics are computed over multiple data partitions, increasing the likelihood that the model will generate reliable predictions on out-of-sample observations. When we apply temporal K-fold cross-validation, we must ensure that training sets contain only observations that precede those in the corresponding test sets. Maintaining this temporal ordering prevents any future information from influencing the model during training, thus preserving the integrity of the forecasting task. Several strategies exist to maintain this temporal constraint, such as sliding windows and expanding windows. In this study, the expanding window method, \Cref{alg:expanding_window}, is selected since it allows for a progressively larger training set, improving the reliability and stability of the resulting forecasts \cite{otext}\cite{hewamalage2023forecast}.

\begin{algorithm}

\SetAlgoLined
\KwIn{$\*X$ with $\{\*x_n\}_{n=1}^N$, indexed by time $n$, initial training size $t$, folds $K$, model $f_*$, error metric $M_\epsilon$}
\KwOut{Cross-validated error}
Let $h \gets \left\lfloor \frac{N - t}{K} \right\rfloor$\;  
Initialize $\text{CV}_{score} \gets 0$\;

\For{$k = 1$ \KwTo $K$}{
    Define training set: $\*{T}_{k} \gets \{ \*x_1, \dots, \*x_{t + (k-1)h} \}$\;
    
    Define validation set: $\*{V}_{k} \gets \{ \*x_{t + (k-1)h + 1}, \dots, \*x_{t + kh} \}$\;
    
    Train the model $f_k$ on $\*T_k$\;
    
    Compute predictions: $\hat{\*{Y}}_{k} \gets f_k(\*V_k)$\;
    
    Let true values: $\*Y_k \gets \text{targets corresponding to } \*V_k$\;
    
    Compute fold error: 
    $\epsilon_{k} \gets M_\epsilon(\*Y_k,\hat{\*Y}_k)$\;
    
    Update: $\text{CV}_{score} \gets \text{CV}_{score} + \epsilon_{k}$\;
}
\Return{$\text{CV}_{score} / K$}\;
\caption{Expanding Window Cross-Validation}
\label{alg:expanding_window}
\end{algorithm}

Furthermore, as addressed in \cite{VUGS_AND_BUCH_TO_THE_TOP}, we employ the optimization framework Optuna, a library specialized for advanced hyperparameter search \cite{akiba2019optuna}.
\section{Experimental setup}
\label{sec:experiments}
The following section delineates the experimental configuration for quantifying the estimated AEP loss, specifically Experiment 1 and Experiment 2. Building on our work from \cite{VUGS_AND_BUCH_TO_THE_TOP}, we implement the same Experiment 2 in the NB-modelling framework to utilize the newly cleaned data for better identification of the substitution year and select a period of sequential identified stable years. 
For Experiment 1, we employ the method of \cite{Byrne2020}, \cite{Astolfi2021V52Aging}, and \cite{Astolfi2022}, where we pick a training year for model training, a reference year and target years for predictions and drift quantification. For Experiment 2, we include all the aforementioned years used in Experiment 1. We train on the entire data set and analyze the drift as a model response by its predictions on a synthetic version of the training data. Experiment 2 is normally known as a sensitivity analysis in the ML regime. Different from Experiment 1 in our work \cite{VUGS_AND_BUCH_TO_THE_TOP}, we now measure the drift as a direct effect of the learned NB model by considering its output residual distributions in subsequent years. Furthermore, the stability analysis delineated in \Cref{subsec:stability_analysis} allows for the automatic selection of train and testing periods as part of the innovative NB framework, in contrast to the methods employed in \cite{VUGS_AND_BUCH_TO_THE_TOP}, where several models were trained in a trial-error fashion to identify usable NB years. Moreover, we employ several model architectures found in the NB research domain to provide a more nuanced view of the AEP loss quantification. Therefore, this section includes the model justifications and their hyperparameter search spaces. Furthermore, we scale the explanatory variables with \Cref{eq:robust_scaler}.

\subsection{Experiment 1.}
\label{subsec:exp1}
The purpose of Experiment 1 is to measure any direct deviation from a WT's expected behavior created under the NB period. The implication is similar to the CM framework by \cite{meyer2021multi}; A model trained on WTs' NB period should exhibit changed residual behavior if a drift occurs in the learned feature relationships. To quantify the change in a meaningful way, we employ the method of \cite{Byrne2020}, \cite{Astolfi2021V52Aging}, and \cite{Astolfi2022}. By identifying periods of stable years, with a minimum sequence of three years, we split the data on a yearly basis as the following:
\begin{itemize}
    \item NB year, the earliest year in the sequence selected for behavioral representation and model training.
    \item Reference year $r$, directly following the NB year.
    \item Target years $t$, subsequently following $r$ and ordered sequentially by time.
\end{itemize}
We train a model on the NB year, and for each subsequent year in the period, we calculate a drift score using the corresponding residuals and actual values on \verb|GridPower|

\begin{equation}
\label{eq:delta}
\Delta=
100
\sum_n^N
\dfrac{\*r_n,}{ \*y_n}. 
\end{equation}
Then, by using the drift score $\Delta_r$ from the reference year, we compute an estimated drift for any of the following target years

\begin{equation}
\label{eq:drift_compare}
\delta_t = \Delta_t-\Delta_r.
\end{equation}

\begin{figure}
    \centering
    \includegraphics[width=1\linewidth]{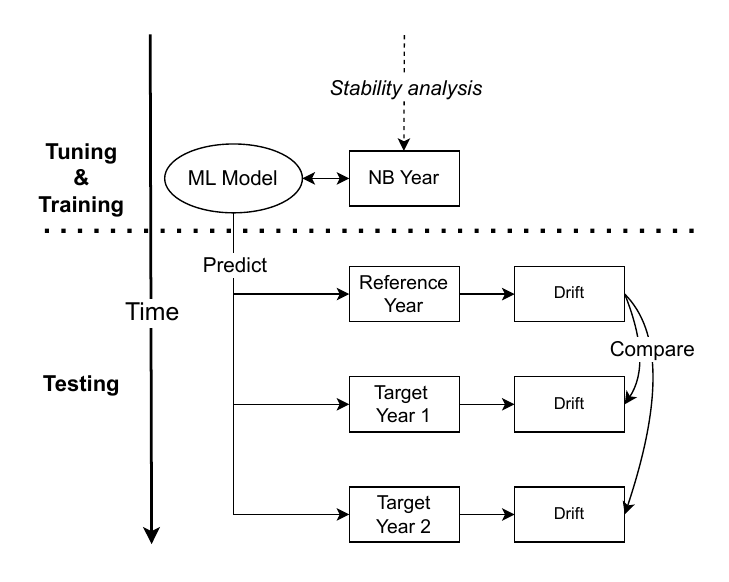}
    \caption{Experiment 1: ML flowchart, modified from \cite{VUGS_AND_BUCH_TO_THE_TOP}}
    \label{fig:exp_1_flow}
\end{figure}
The entire process is visualized on \Cref{fig:exp_1_flow} and operates as a direct extension to the NB framework. By clearly defining criteria for the stable periods and employing our data-splitting strategy, we automate the entire process of data selection, training, and AEP loss quantification. From empirical analysis and to provide a comparative period, we constrain the NB year to be either 2018 or 2019.

\subsection{Experiment 2.}
\label{subsec:exp2}
Because Experiment 1 is a direct response from the NB model on to subsequent years, we include Experiment 2 from our earlier work \cite{VUGS_AND_BUCH_TO_THE_TOP} for a more nuanced view of the potential underlying feature drift realized by time. The following explanation is a modified version from \cite{VUGS_AND_BUCH_TO_THE_TOP}. To incorporate the NB year in the AEP loss quantification, Experiment 2 contains the same stable periods used for Experiment 1, encompassing at least three subsequent years for individual WTs. For each turbine, a group of models is trained on the stable period, and the year of each observation is included. By doing so, we can perform a sensitivity analysis for the models' responses to the yearly operating conditions, subjugated to the learned effect of the NB year.

We construct a synthetic dataset by permuting the original records to examine how the models' performances evolve for each subsequent year following the chosen NB year. In generating this synthetic dataset, observations are replicated in proportion to the remaining years following the NB year. For each replication, the values for \verb|Year| are substituted with the NB year, leaving the dataset otherwise identical to the original. The distinctive yearly effect of the NB year is then quantified by allowing the turbine-specific models to predict \verb|GridPower| on this synthetic dataset. This process yields an estimated AEP drift, expressed as a percentage relative to the chosen NB year, as defined in \Cref{eq:delta}.

\begin{table}[htbp]
\centering
\caption{Hyperparameter Search Space. A tuple $(\text{a},\text{b})$ indicates a continuous value range, and a list $[\text{a},\text{b}]$ denotes categorical values.}
\label{tab:parameter_space}
\begin{tabular}{|l|l|l|}
\hline
\textbf{Algorithm} & \textbf{Parameter}         & \textbf{Search space}          \\ \hline\hline
Multi-layer Perceptron       & learning\_rate             & (0.01, 0.1)                    \\ \hline\hline
Random Forest        & n\_estimators              & (10, 1000)              \\ \cline{2-3}
                   & min\_samples\_leaf         & (1, 100)                \\ \cline{2-3}
                   & min\_samples\_split        & (2, 100)                \\ \hline\hline
XGBoost       & n\_estimators              & (10, 1000)              \\ \cline{2-3}
                   & min\_child\_weight         & (1, 100)                \\ \cline{2-3}
                   & max\_depth                 & (1, 10)                 \\ \cline{2-3}
                   & learning\_rate             & (0.001, 1)              \\ \cline{2-3}
                   & min\_split\_loss           & (0, 10)                 \\ \cline{2-3}
                   & colsample\_bytree          & (0.1, 1)                \\ \cline{2-3}
                   & lambda                     & (0, 10)                 \\ \cline{2-3}
                   & alpha                      & (0, 10)                 \\ \cline{2-3}
                   & n\_jobs                    & [1]                     \\ \hline\hline
K-Nearest Neighbor      & n\_neighbors               & (1, 1000)               \\ \cline{2-3}
                   & p                          & [1, 2]                  \\ \hline
\end{tabular}
\end{table}

\subsection{Hyperparameter Tuning and Model Selection}
The models included in this study are selected based on prior reports of their performances within the research domain. For instance, \cite{Janssens2016} reports a generally solid performance of the Random Forest algorithm across uni and multivariate PC modeling. Furthermore, \cite{Janssens2016} also reports that the KNN model showcases the overall best performance on the univariate PC modeling. \cite{mathew2022estimation} reports a good performance using the MLP, even compared to other tested models in their study. Lastly, we reported in \cite{VUGS_AND_BUCH_TO_THE_TOP} how XGBoost exhibits solid modeling potential of WT NB. None of the before-mentioned studies reports their model MAPE, making it difficult to compare models across studies directly. Furthermore, what constitutes NB varies from study to study. \cite{mathew2022estimation} reports using approximately the same parts of the PC we employ in this study and in \cite{VUGS_AND_BUCH_TO_THE_TOP}. However, that is not the case for \cite{Janssens2016}, which includes the rated power, a region generally exhibiting little variance and noise. Nevertheless, we include these models to provide a nuanced perspective of the results.

To ensure the quality of the selected models during testing, they undergo hyperparameter selection through the framework briefly described in \Cref{subsec:hyper_teori}. Every model, for every WT, is tuned using Optuna with the expanding window method (\Cref{alg:expanding_window}), with five folds and $50$ trials. We include the parameter search spaces in \Cref{tab:parameter_space} for reproducibility, showcasing model complexity. For the MLP, it should be noted that we use the architecture of \cite{mathew2022estimation}, with input features $D$, giving a layer flow with a number of neurons 
$
D\rightarrow16\rightarrow32\rightarrow1.
$ 
We use the Adam optimizer and a validation set for early stopping. The validation set contains the last $20\%$ of the training data. The early stopping criteria are based directly on the target value. If there is no improvement of at least $1\%$ validation error of the max target value over $15$ epochs, we stop the training and use the model's state that meets the set criteria. If the reader is interested in the hyperparameters of \Cref{tab:parameter_space} effects on their respective models, we refer to their library documentations:
\begin{itemize}
    \item \verb|sklearn| \cite{scikit-learn-docs}: K-Nearest Neighbor, Random Forest.
    \item \verb|Pytorch| \cite{pytorch2025docs}: Multi-layer Perceptron.
    \item \verb|XGBoost| \cite{XGBoost_doc}: Xgboost.
\end{itemize}

The hyperparameter selection and re-training of the models constitutes a severe computational hurdle; for every WT, using \Cref{alg:expanding_window}, we train a candidate model $250$ times per experiment, which is computationally expensive. We, therefore, tune, train, and test the models in parallel, utilizing the python \verb|multiprocessing| module. Therefore, we include the \verb|n_jobs| parameter for XGBoost, mitigating computational bottlenecks and overhead, and the MLP is trained on the CPU rather than the CUDA framework.

\begin{table*}[t]
     \caption{Data remaining by applying the data preprocessing methods, reported in thousands. \% Hard-filter and \% NB-filter describes overall removed data by the methods, While Hard-filter Data and NB-filter Data is the remaining data after the applied methods.  
    }
    \label{tab:data-pipeline-result}
    \centering
    \begin{tabular}{|c|c||c|c||c|c||c|}
    \hline
    {\textbf{WF}} &  {\textbf{Raw Data}} & {\textbf{\% Hard-filter}} &  {\textbf{Hard-filter Data}} & {\textbf{NB-filter Data}} & {\textbf{\% NB-filter}} & {\textbf{\% Remaining}} \\ \hline
    1  & 3993 & 73 & 1094 & 1004  & 2 &\textbf{25} \\ \hline
    2  & 3608 & 70 & 1082 & 989   & 3 &\textbf{27} \\ \hline
    3  & 1742 & 64 & 625  & 601   & 1 &\textbf{35} \\ \hline
    5  & 2716 & 67 & 905  & 850   & 2 &\textbf{31} \\ \hline
    6  & 2558 & 66 & 867  & 670   & 8 &\textbf{26} \\ \hline
    7  & 2323 & 67 & 776  & 672   & 4 &\textbf{29} \\ \hline
    8  & 4427 & 60 & 1787 & 1477  & 7 &\textbf{33} \\ \hline
    9  & 3026 & 59 & 1236 & 1088  & 5 &\textbf{36} \\ \hline
    10 & 3550 & 64 & 1263 & 1184  & 2 &\textbf{33} \\ \hline
    11 & 4136 & 63 & 1510 & 1208  & 7 &\textbf{29} \\ \hline
    12 & 3570 & 63 & 1338 & 1146  & 5 &\textbf{32} \\ \hline
    13 & 3552 & 60 & 1406 & 1305  & 3 &\textbf{37} \\ \hline
    14 & 4606 & 73 & 1223 & 1129  & 2 &\textbf{25} \\ \hline
    15 & 4599 & 64 & 1666 & 1553  & 3 &\textbf{34} \\ \hline
    16 & 4579 & 61 & 1787 & 1593  & 4 &\textbf{35} \\ \hline
    \end{tabular}
\end{table*}

\section{Results}\label{sec:results}
In this section, we present the results obtained from performing the experiments delineated in \Cref{sec:experiments} and present the effects of the data preprocessing. Following the application of the data preprocessing workflow and the criteria for stable periods, we are left with a total of $35$ WTs out of $117$ for Experiments 1 and 2, with up to four target years for some turbines. However, with shallow model testing and initial preliminary analysis, it became clear that even with the excessive data cleaning, the models still experienced behavior similar to our work in \cite{VUGS_AND_BUCH_TO_THE_TOP}. That is, using the ML framework and \Cref{eq:drift_compare}, the WTs show an increase in energy efficiency, contradicting the literature. In consultancy with Vestas, we believe this issue arises from multiple re-adjustments on the blade load sensors. We previously believed, together with our contact point at Vestas, that these adjustments happen at low frequencies and were one of the motivating factors for the initial use of stable period selection. We include these results in this section to provide evidence of this phenomenon and further discuss them in \Cref{sec:discussion} with illustrations. We also include the PC for experimentation to circumvent the effect of sensor adjustments. In \cite{meyer2020data}, the air temperature is utilized as a partial proxy for air pressure in PC modeling. We similarly include this variable, \verb|AmbTemp|, allowing the models to smooth the WS binning on the PC. We distinguish the two variables set $\nabla$ with \textit{PC}, containing the \verb|WindSpeed| and \verb|AmbTemp|, and \textit{All}, containing all explainable variables except \verb|WSE|. From the $35$ WTs included in the experiment, we select and present two different WTs, $4$ and $80$. WT $4$ represents the vast majority of observed behavior among the $35$ WTs, whereas WT $80$ represents a small subset, exhibiting behavior similar to the turbines isolated in \cite{VUGS_AND_BUCH_TO_THE_TOP}. 
\begin{figure}[h]
    
    \centering
    \includegraphics[width=1\linewidth]{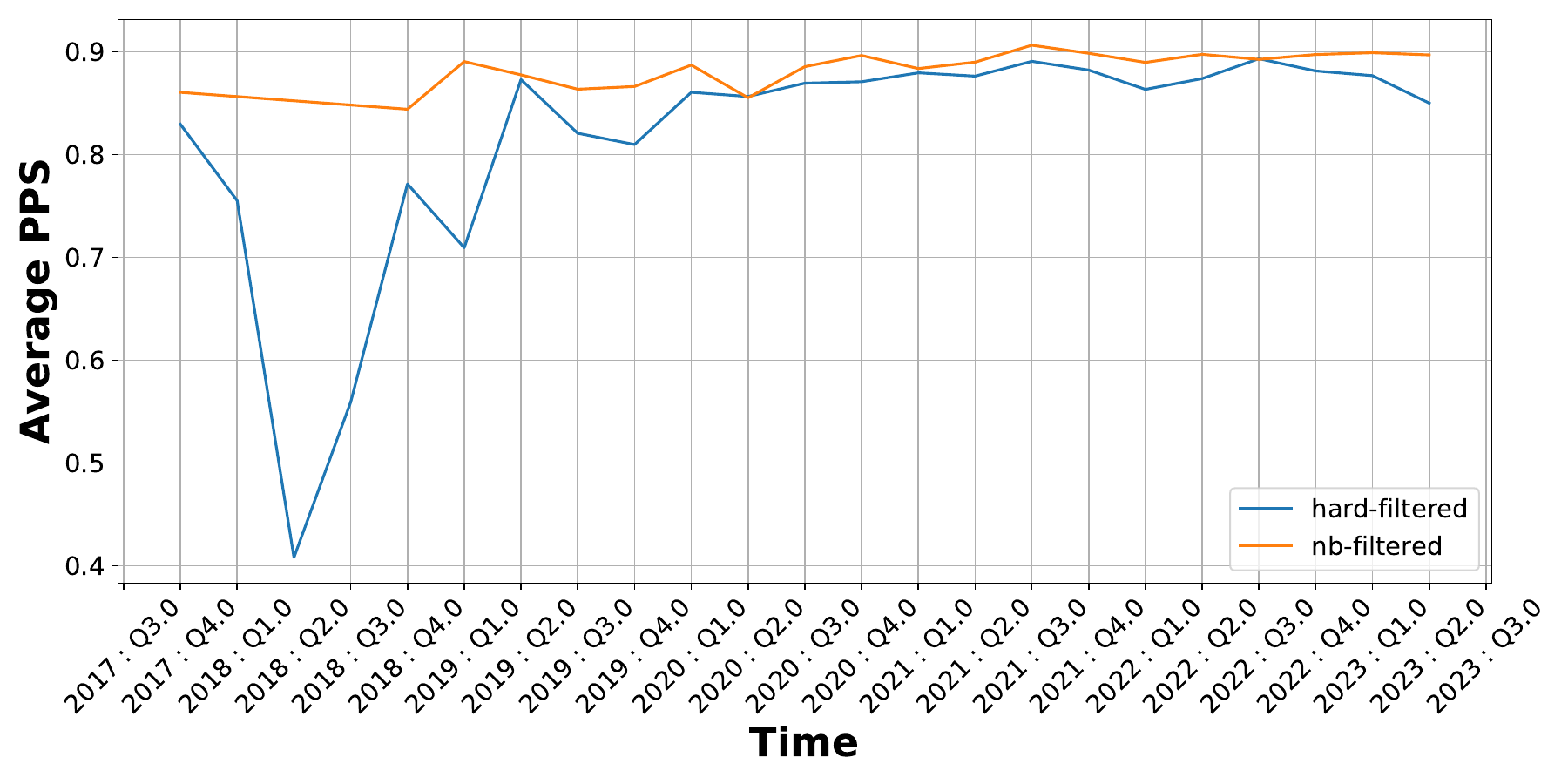}
    \caption{Average measured PPS between NB-filtered and hard-filtered data for WT 31.}
    \label{fig:avg_pps_NB_vs_HF}
\end{figure}
\subsection{Preprocessing}
The preprocessing results can be observed on \Cref{tab:data-pipeline-result}. On average, we keep around $31\%$ of the data per WF, with the minimum remaining data being $25\%$ of the original and the maximum $37\%$. We see the hard-filters being the main cause of the data removed, with NB-filters making up only a fraction of the total data removal. The hard-filters are justified by the resulting operating state, as delineated by \Cref{tbl:hard-filter}, and to showcase the justification of the NB-filters, we showcase a WT from the remaining 35 turbines, WT 31, \Cref{fig:avg_pps_NB_vs_HF} as an example. On \Cref{fig:avg_pps_NB_vs_HF}, it can be observed how we not only increase the overall PPS but also directly expand a potential window for possible, stable period modeling by pinpointing and removing the data that decreases the PPS. Because we measure the data quality directly as a measure of the PPS, the NB-filters are justified and provide a promising data-cleaning framework.

\subsection{Experiment 1}
When we employ the feature set \textit{All}, we can observe a generally low MAPE on \Cref{tab:exp_1_4}. It is evident that WT $4$ exhibits clear improvement through its operation history, characterized by its positive drift compared to the reference year. All the models primarily agree on a stable positive drift from 2020 through 2022, with three models quantifying a sudden drop in performance in 2023. We see a generally good MAPE across all models, with the MLP showcasing the worst MAPE in the reference year. This is the predominant case for the MLP across the conducted experiments. When applying the feature set \textit{PC}, the quantified drift of the WT changes completely as observed on \Cref{tab:exp_1_4}. The performance deterioration of the PC has been subtle but consistent throughout the last years of the operation history, with the RF, XGBoost, and KNN showcasing similar results across the line. By using the PC as a reference, we trade the precision of the feature set \textit{All} to facilitate the quantification of performance degradation, as evidenced by the overall rise of MAPE between the feature sets.   


\begin{table}[!ht]
\centering
\caption{Experiment 1 drift results: WT 4, with MAPE for the reference year.}
\label{tab:exp_1_4}
\begin{tabular}{|c|c|c|c|c|c|c|}
\hline
\textbf{$\nabla$} & \textbf{Model} & \textbf{MAPE $r$} & \textbf{2020} & \textbf{2021} & \textbf{2022} & \textbf{2023} \\
\hline\hline
\multirow{4}{*}{All} 
    & RF      & 4.9 & 7.2 & 7.1 & 7.8 & 7.7 \\
\cline{2-7}
    & XGBoost & 4.8 & 5.9 & 6.1 & 6.2 & 4.9 \\
\cline{2-7}
    & KNN     & 5.1 & 2.8 & 2.8 & 3.0 & 1.4 \\
\cline{2-7}
    & MLP     & 7.0 & 2.9 & 2.9 & 3.0 & 1.9 \\
\hline\hline
\multirow{4}{*}{PC}

    & RF      & 6.6 & -0.3 & -0.7 & -0.5 & -1.4 \\
\cline{2-7}
    & XGBoost & 6.7 & -0.1 & -0.7 & -0.4 & -1.5 \\
\cline{2-7}
    & KNN     & 7.5 & -0.1 & -0.8 & -0.3 & -1.4 \\
\cline{2-7}
     & MLP     & 8.4 & 0.2  & -0.9 & 0.1  & -0.5 \\
\hline
\end{tabular}
\end{table}



\begin{table}[h]
\centering
\caption{Experiment 1 drift results: WT 80, with MAPE for the reference year.}
\label{tab:exp_1_6}
\begin{tabular}{|c|c|c|c|c|c|c|}
\hline
\textbf{$\nabla$} & \textbf{Model} & \textbf{MAPE $r$} & \textbf{2020} & \textbf{2021} & \textbf{2022} & \textbf{2023} \\
\hline\hline
\multirow{4}{*}{All}
    & RF      & 5.7  & -3.7 & -11.2 & -8.7 & -6.7 \\
\cline{2-7}
    & XGBoost & 5.9  & -3.1 & -10.8 & -7.6 & -6.1 \\
\cline{2-7}
    & KNN     & 5.9  & -2.8 & -11.1 & -7.4 & -6.3 \\
\cline{2-7}
    & MLP     & 13 & -2 & -13.7 & -6.1 & -4.6 \\
\hline\hline
\multirow{4}{*}{PC}

    & RF      & 6.9  & -2.9 & -11.2 & -10.8 & -8.6 \\
\cline{2-7}
    & XGBoost & 7.2  & -2.9 & -11.2 & -10.9 & -8.5 \\
\cline{2-7}
    & KNN     & 7.1  & -2.8 & -11.1 & -10.7 & -8.4 \\
\cline{2-7}
        & MLP     & 8.6  & -2.4 & -11.9 & -10.4 & -7.7 \\
\hline
\end{tabular}
\end{table}

Next, we consider the Experiment 1 results for WT 80, picked for its peculiar behavior. For the feature set \textit{All}, we can observe a reasonable MAPE for three of the models with a negative drift. Specifically, 2021 showcases an extreme negative drift across all the models, signaling some potential underlying health issues. However, the change in performance was not flagged as having a low PPS; thus, it is still included among the turbines considered stable. The following years exhibit a positive drift away from 2021, indicating that adjustments are still being made but are insufficient to offset the drop completely in 2021. When the \textit{PC} feature set is employed, we see similar behavior for 2020 and 2021, with KNN and XGBoost having identical results to their \textit{All} counterparts. Furthermore, 2022 and 2023 exhibit a negative drift compared to the reference year. However, compared to 2021, they have a positive drift, supporting the issues highlighted by the \textit{All} feature set.     

When we look at the PC over the $35$ WTs, $24$ exhibits a clear decline in performance, seven improve, and four show no significant improvement nor decline in performance. 
\begin{table}[h]
\centering
\caption{Experiment 2 drift results: WT 4, with MAPE for the reference year from Experiment 1.}
\label{tab:exp_2_4}
\begin{tabular}{|c|c|c|c|c|c|c|}
\hline
\textbf{$\nabla$} & \textbf{Model} & \textbf{MAPE $r$} & \textbf{2020} & \textbf{2021} & \textbf{2022} & \textbf{2023} \\
\hline\hline
\multirow{4}{*}{All} 
    & RF      & 1.6 & 0 & 0.2  & -0.5 & -2\\
\cline{2-7}
    & XGBoost    & 2.8 & 0.7   & 1.1  & 0  & -2.3 \\
\cline{2-7}
    & KNN    & 4.7 & 2.9   & 3   & 3.1  & 1.5 \\
\cline{2-7}
    & MLP     & 7 & -1.4  & -1.7  & -1.2  & -2.4 \\
\hline\hline
\multirow{4}{*}{PC}
    & RF      & 6.7  & -0.1 & -0.6 & -0.4 & -1.3 \\
\cline{2-7}
    & XGBoost & 6.8  & -0.1 & -0.6 & -0.4 & -1.3 \\
\cline{2-7}
    & KNN     & 6.7  & -0.2 & -0.4 & -0.1 & -1.1 \\
\cline{2-7}
    & MLP     & 11  & -4.8 & -5.8 & -5.2 & -7.2 \\
\hline
\end{tabular}
\end{table}
\begin{table}[h]
\centering
\caption{Experiment 2 drift results: WT 80, with MAPE for the reference year from Experiment 1.}
\label{tab:exp_2_80}
\begin{tabular}{|c|c|c|c|c|c|c|}
\hline
\textbf{$\nabla$} & \textbf{Model} & \textbf{MAPE $r$} & \textbf{2020} & \textbf{2021} & \textbf{2022} & \textbf{2023} \\
\hline\hline
\multirow{4}{*}{All}

    & RF      & 3.2 & -3.1 & -10.9  & -8.1 & -6.2\\
\cline{2-7}
    & XGBoost    & 5.2 & -3   & -11.2  & -9.3  & -7.7 \\
\cline{2-7}
    & KNN    & 5.4 & -3.1   & -11.6   & -7.7  & -6.5 \\
\cline{2-7}
    & MLP     & 10 & 5.9  & -2.2  & 2  & 2.6 \\
\hline\hline

\multirow{4}{*}{PC}

    & RF      & 6.4  & -2.7 & -11 & -10.6 & -8.3 \\
\cline{2-7}
    & XGBoost & 7.2  & -2.6 & -10.9 & -10.8 & -9.1 \\
\cline{2-7}
    & KNN     & 6.5  & -2.8 & -11 & -10.6 & -8.4 \\
\cline{2-7}
    & MLP     & 13.2  & 9.6 & 2.8 & 2.1 & 4.1 \\
\hline
\end{tabular}
\end{table}

\subsection{Experiment 2}
For Experiment 2, we include a calculated MAPE from the same reference year used in Experiment 1 as a point of reference. It is, however, important to emphasize that Experiment 2 utilizes the entire stable period as a training set. Therefore, the MAPE is an in-sample error, and a reduction in MAPE from Experiment 1 is expected. For WT 4 with feature set \textit{All}, \Cref{tab:exp_2_4}, we see similar results from what we observed in \cite{VUGS_AND_BUCH_TO_THE_TOP} for the XGBoost and the RF models; the initial years, 2020 and 2021 showcase no or slight positive effect on AEP, manifesting into a negative drift in 2023. The KNN and MLP exhibit widely different behavior compared to the tree models, with the MLP explicitly quantifying a negative drift. However, the MLP has not seen a decrease in MAPE compared to Experiment 1, and we should question the validity of these results. When we consider the \textit{PC} features, something interesting appears: The three models, RF, XGBoost, and KNN, produce almost identical results to Experiment 1. This find is particularly relevant for several reasons: 
\begin{enumerate}[label=\roman*.]
    \item It signals that the PC has stable readings; the only change is isolated to a yearly effect. 
    \item It indicates the yearly effect, quantified under a sensitivity analysis and agreeing with the NB framework, underscores the direct effect of time on the AEP.
    \item The sensitivity analysis validates the legitimacy of the NB and CM approaches we and other studies have applied.
    \item Utilizing the NB framework is significantly cheaper computationally, requiring only a fraction of the training data needed compared to the sensitivity analysis.  
\end{enumerate}
If we look at \Cref{tab:exp_2_80} for WT 80 from Experiment 2 and compare it to Experiment 1, we still see overall agreement between the experiments and the feature sets $\nabla$. This indicates that drift between the explainable variables may not have occurred to an extreme degree during the identified stable period. However, it is important to outline that the behavior of WT 80 is a minority among the 35 remaining turbines and is explored further in \Cref{sec:discussion}.
\section{Discussion}\label{sec:discussion}
\begin{figure}[ht]
    \centering
    \includegraphics[width=1\linewidth]{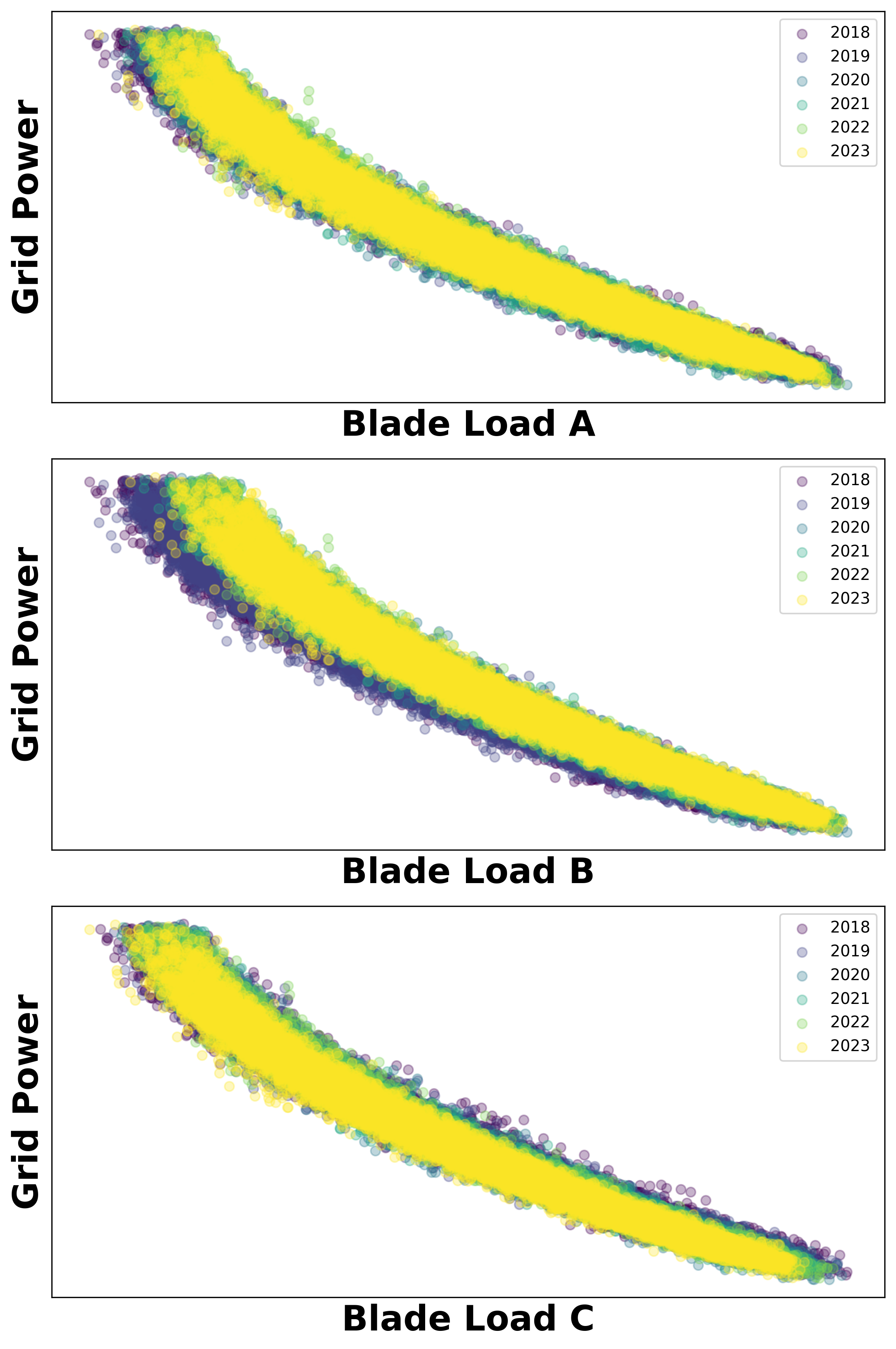}
    \caption{Wind Farm Id 1 wind turbine 4 - Yearly drift BladeLoad for selected stable years}
    \label{fig:p1t4-drift-bl}
\end{figure}

In \Cref{sec:results}, we observe patterns similar to findings from our previous study \cite{VUGS_AND_BUCH_TO_THE_TOP}, wherein certain turbines exhibit an apparent increase in performance over time. This trend persists across various model configurations as shown in \Cref{tab:exp_1_4}. Specifically, for models that include the full set of explanatory variables, the calculated drift values are predominantly positive, suggesting an improvement in performance over time. As discussed in \Cref{sec:results}, this trend can be attributed to periodic recalibrations performed by Vestas on the SCADA system. These recalibrations are illustrated in \Cref{fig:p1t4-drift-bl}, which shows consistent behavior over time for \verb|BladeLoadA| and \verb|BladeLoadC|, whereas \verb|BladeLoadB| displays a noticeable shift in its relationship with \verb|GridPower| beginning in 2021. Post 2021, \verb|BladeLoadB| exhibits lower measured loads for equivalent power output levels, implying that less aerodynamic load is required to achieve the same energy production. This shift suggests that the turbine can produce similar output at reduced wind speeds relative to earlier periods due to the recalibration.

\begin{figure}[ht]
    \centering
    \includegraphics[width=1\linewidth]{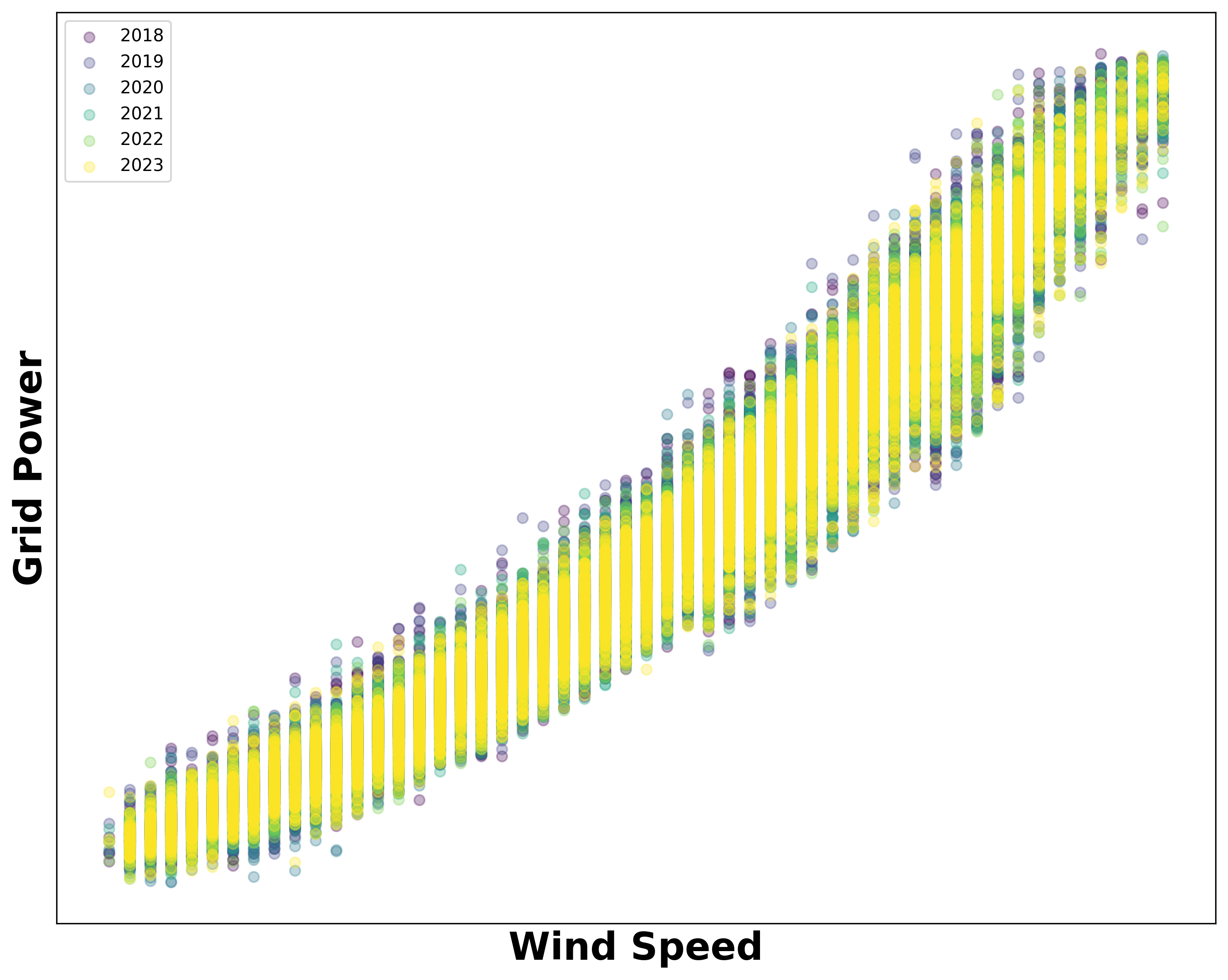}
    \caption{Wind Farm Id 1 wind turbine 4 - Yearly drift PC for selected stable years}
    \label{fig:p1t4-drift-pc}
\end{figure}

However, a contrasting trend emerges when examining the same models in \Cref{tab:exp_1_4}, but restricting the \textit{PC} feature set, which consists only of \verb|WindSpeed| and \verb|AmbTemp|. In this case, most models exhibit a small negative drift, indicating a slight degradation in performance over time. Additionally, we observe an increase in the MAPE, which is expected when excluding variables that contribute to explaining variability in the target variable. As shown in \Cref{fig:p1t4-drift-pc}, no recalibration effects are apparent for the PC, which is consistent with Vestas’ practice of rarely adjusting the wind speed sensor.

\begin{figure}[ht]
    \centering
    \includegraphics[width=1\linewidth]{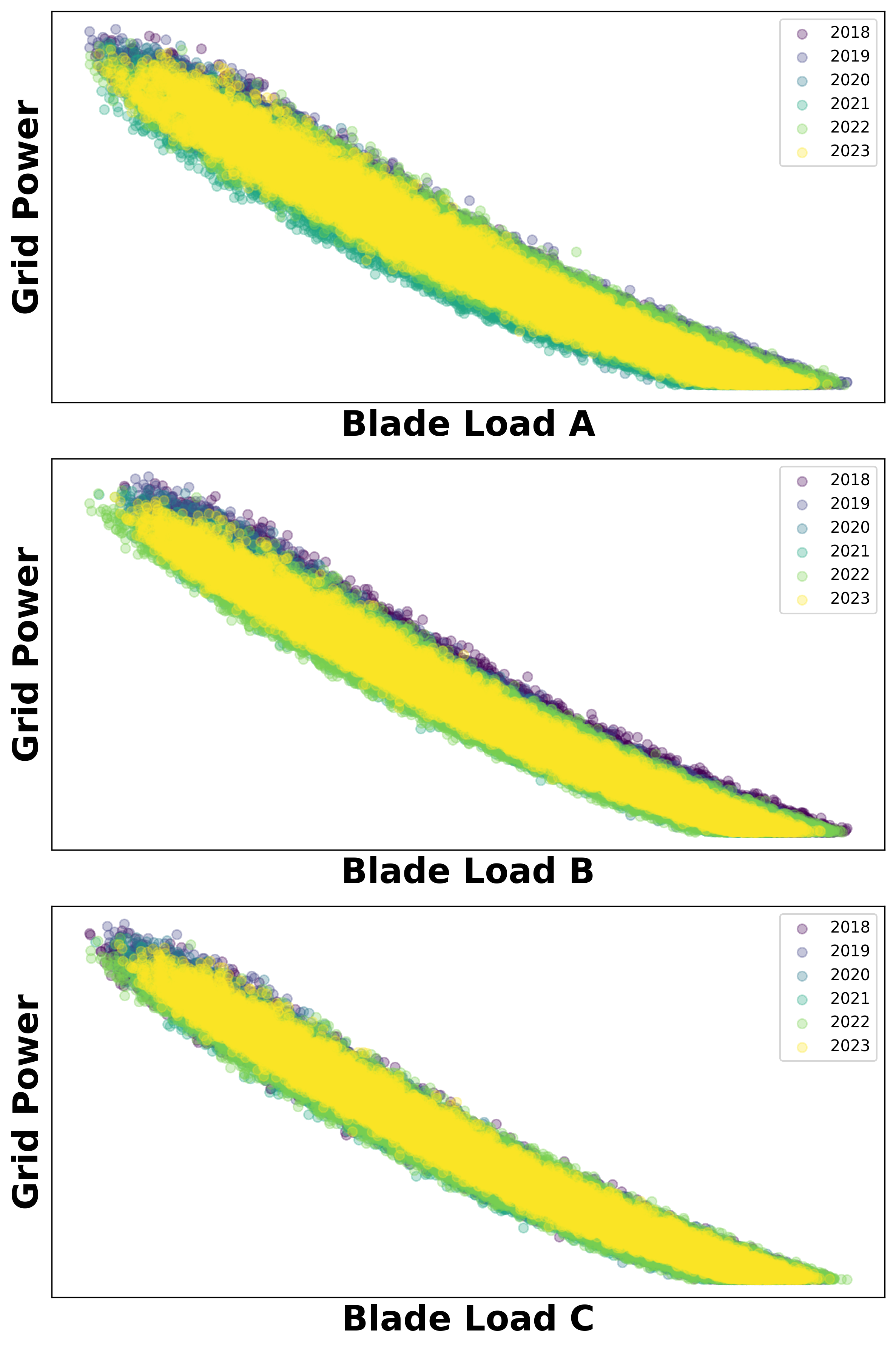}
    \caption{Wind Farm Id 12 wind turbine 80 - Yearly drift BladeLoad for selected stable years}
    \label{fig:p12t80-drift-bl-stable}
\end{figure}

\begin{figure}[ht]
    \centering
    \includegraphics[width=1\linewidth]{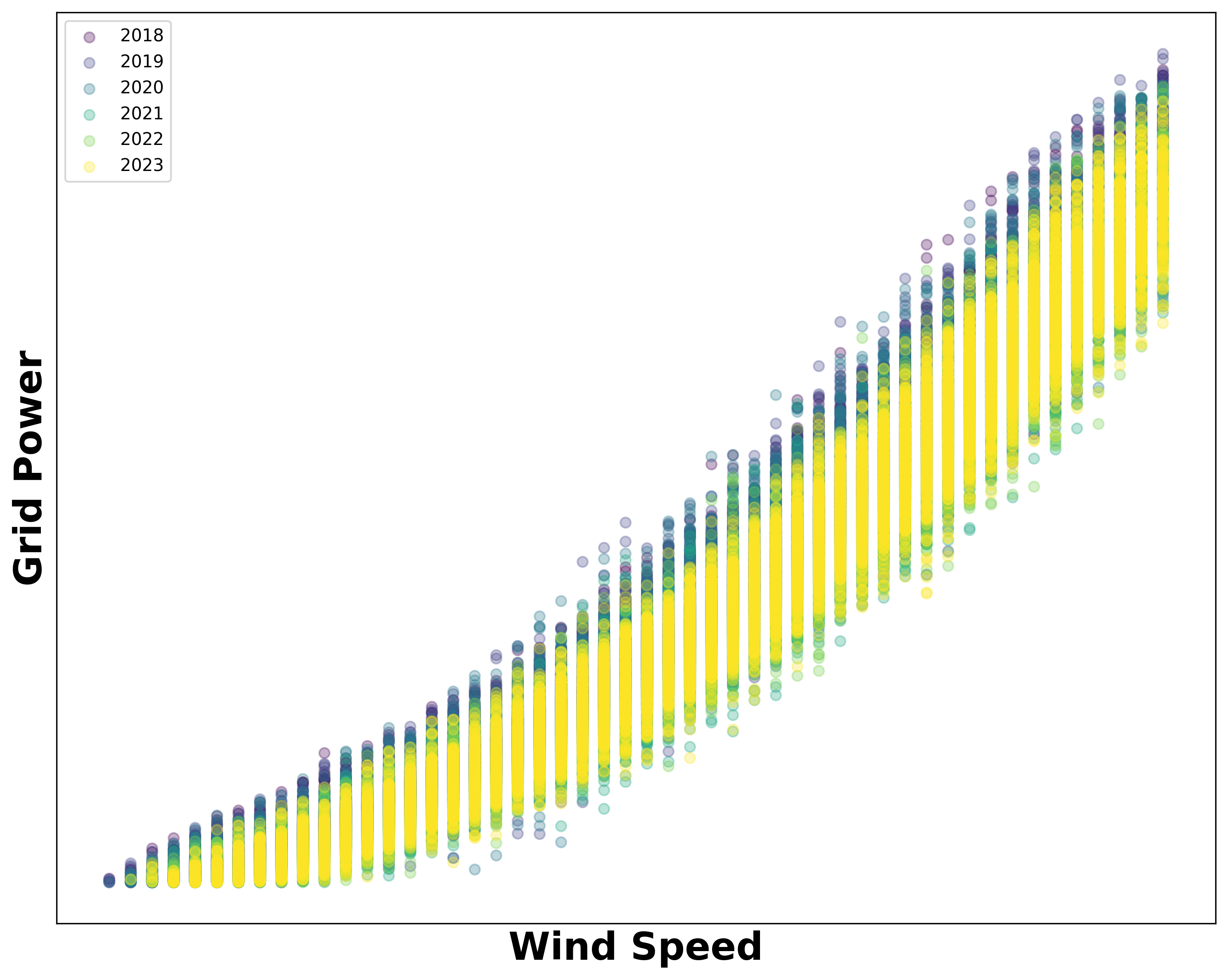}
    \caption{Wind Farm Id 12 wind turbine 80 - Yearly drift PC for selected stable years}
    \label{fig:p12t80-drift-pc-stable}
\end{figure}

The results presented in \Cref{tab:exp_1_6} align with the findings reported in the literature cited throughout this study. Specifically, we observe a significant negative drift, indicating turbine behavior degradation. This makes the corresponding WT a strong candidate for further investigation by Vestas to validate its on-site operational condition. Such validation would confirm the observed performance decline and demonstrate the practical utility of the proposed framework as a CM tool. By identifying turbines exhibiting significant performance drift, the framework supports Vestas’ efforts in data-driven decision-making for on-site maintenance. The negative drift identified in \Cref{tab:exp_1_6} is further validated \Cref{fig:p12t80-drift-bl-stable}, which shows no evidence of recalibration in the blade load signals. Similarly, observations hold for the PC in \Cref{fig:p12t80-drift-pc-stable}, where the presumed absence of recalibration reinforces the validity of the detected performance decline.

\begin{figure}[ht]
    \centering
    \includegraphics[width=1\linewidth]{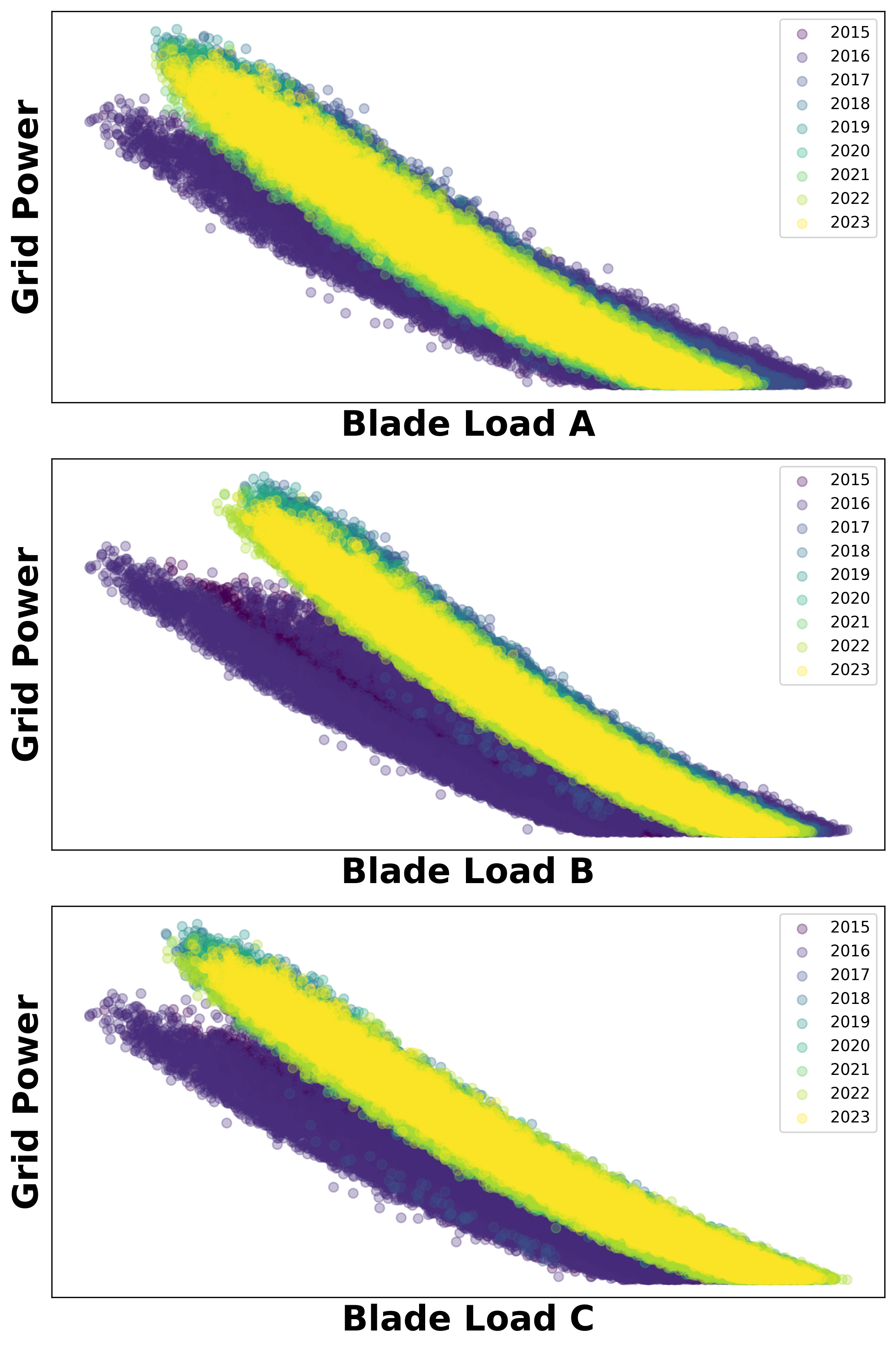}
    \caption{Wind Farm Id 12 wind turbine 80 - Yearly drift BladeLoad}
    \label{fig:p12t80-drift-bl}
\end{figure}

As outlined in \Cref{sec:results}, the initial dataset comprises 117 WTs. After applying the complete data preprocessing pipeline, including the hard-filters, NB-filters, and the stability-based selection methodology, we reduce the dataset to 35 turbines. \Cref{fig:p12t80-drift-bl-stable}, represents the result of our data preprocessing pipeline for multiple WTs and displays the blade load signals and corresponding energy output for turbines during their identified stable operational years. In contrast, excluding the stability framework and applying only the hard-filters and NB-filters yields the results shown in \Cref{fig:p12t80-drift-bl}. This reveals substantial changes across all blade load signals, specifically in the years preceding 2018, indicating that sensor recalibrations likely occurred during that period. Our previous study \cite{VUGS_AND_BUCH_TO_THE_TOP} does not capture this level of detail, as it relies on manually selected training, reference, and testing periods, an approach that is feasible only when analyzing a small number of turbines. The current results highlight the scalability and flexibility of the proposed framework. By automating the identification of stable operational periods and systematically isolating NB data, the methodology effectively accounts for temporal inconsistencies such as sensor recalibrations. Although not without limitations, the framework marks a novel improvement in preprocessing strategies for data-driven CM in wind energy systems.

\section{Conclusion}\label{sec:conlusion}
In this study, we have developed a general and scalable data-driven approach to quantify the AEP loss manifested through a WT's operational lifespan. Through our extensive and systematic data preprocessing steps involving both hard-filters and NB-filters, we successfully isolate NB SCADA data for further downstream modeling and analysis tasks. Without the event logs from the operational history, it is a non-trivial challenge to identify NB-suitable periods for modeling. However, through the temporally transformed PPS, we demonstrate how it is still possible to build a scalable method to identify and model NB periods with a continuously stable operational history for AEP loss quantification.

The turbine-tailored ML models provided insight into the AEP degradation of individual turbines and provided valuable analytics for future maintenance planning and fundamentals for financial decisions. We presented WT 4, showcasing a slow but steady performance degradation through the quantified period, and WT 80, with a steep decline in performance and overall negative performance drift, when utilizing the \textit{PC} feature set. With the application of several ML models, we showcase turbine-tailored models exhibiting low MAPEs and precise predictions, enabling accurate estimates of the performance degradations through two different experimental approaches. An important discovery was made during this experimental comparison; the PC for both experiments seems to vastly agree on AEP quantifications, signaling that the reliance on the computationally cheaper method employed by Experiment 1 can be sufficient if the underlying variable relationships are only affected by time. We suspect the same results are not agreeable for the variable set we named \textit{All} due to frequent signal recalibrations. In this case, Experiment 2 provides a model-based opportunity to learn the changing relationships under training, and we measure the impact of the yearly variable as a model response.   

The weakness of this study arises directly from the limitations of anonymized SCADA data. With critical information like small signal recalibrations obfuscated, it becomes increasingly harder to identify changes such as temporal degradation from signal adjustments. This is showcased by the experimental setup, where running recalibrations are still present and not discovered during the PPS analysis. Because the temporal PPS is a cross-validated score, it relies on significant variable drifts to flag a period, rendering it blind to smaller recalibrations with the used thresholds. Nevertheless, it is still showcased as working exceptionally well when applied with the PC for downstream ML tasks. 

In conclusion, our study contributes to the literature, providing a scalable framework for accurately quantifying energy production losses and improving management and maintenance decisions in the wind energy sector.
\section{Future work}\label{sec:future-work}
Building upon the contributions outlined in this research, several promising avenues remain to advance data-driven methodologies in WT CM and predictive maintenance. Firstly, refining the unsupervised identification and modeling of healthy operating states remains critical. Exploring advanced clustering and anomaly detection algorithms can enhance the precision of NB models when distinguishing subtle performance deviations that indicate energy loss. In \cite{VUGS_AND_BUCH_TO_THE_TOP}, we previously proposed the use of unsupervised autoencoders for anomaly detection. However, this approach relies on the assumption that the majority of the dataset represents normal operational conditions, an assumption that was not verifiable at the beginning of this project.

Furthermore, autoencoders belong to the class of black-box models, which limits interpretability and makes it difficult to understand the underlying reasons for classifying specific observations as an anomaly. It is showcased by \Cref{tab:data-pipeline-result} that NB-filters exclude only a fraction of the original data, indicating that the majority of the remaining data points after hard-filtering represent NB. Thus, using autoencoders as an NB-filter could increase the refinement of the NB isolation process. 
 
Incorporating more comprehensive external meteorological data can improve model estimations of AEP loss and NB data isolation. Access to detailed meteorological parameters such as precipitation, humidity, and air density can significantly improve the predictive accuracy of CM strategies. Integrating event logs that describe recalibration and maintenance events can improve the identification of NB states with greater precision. These suggestions can help with ML-driven CM solutions' accuracy and reliability, improving their utility and efficiency in real-world deployments.
\section*{Acknowledgements}\label{sec:acknowledgements} 
This study was made possible by Aalborg University (AAU) and Vestas. The dataset was provided by Vestas and is kept private as it contains valuable production information. We personally thank Christian Schilling of Aalborg University for supervising the work developed and presented in this paper, Johnny Nielsen of Vestas for close collaboration and domain guidance, and Thomas Dyhre Nielsen of Aalborg University for ML-related guidance and consultancy.

We leveraged the assistance of ChatGPT \cite{chatgpt} combined with ResearchRabbit \cite{researchrabbit2025} for discovering valuable research sources through deep research and citation tracking. Furthermore, GitHub Copilot \cite{copilot} provided valuable basic code suggestions, expediting the development process substantially. Lastly, for the final proofreading, we utilized Grammarly \cite{grammarly} for grammatical suggestions and reading clarity.
\printbibliography



\appendices
\section{Illustrations of multi-objective K-selection for GMM-components} \label{appendix:k-selection}
\begin{figure}[h]
    \centering
    \includegraphics[width=1\linewidth]{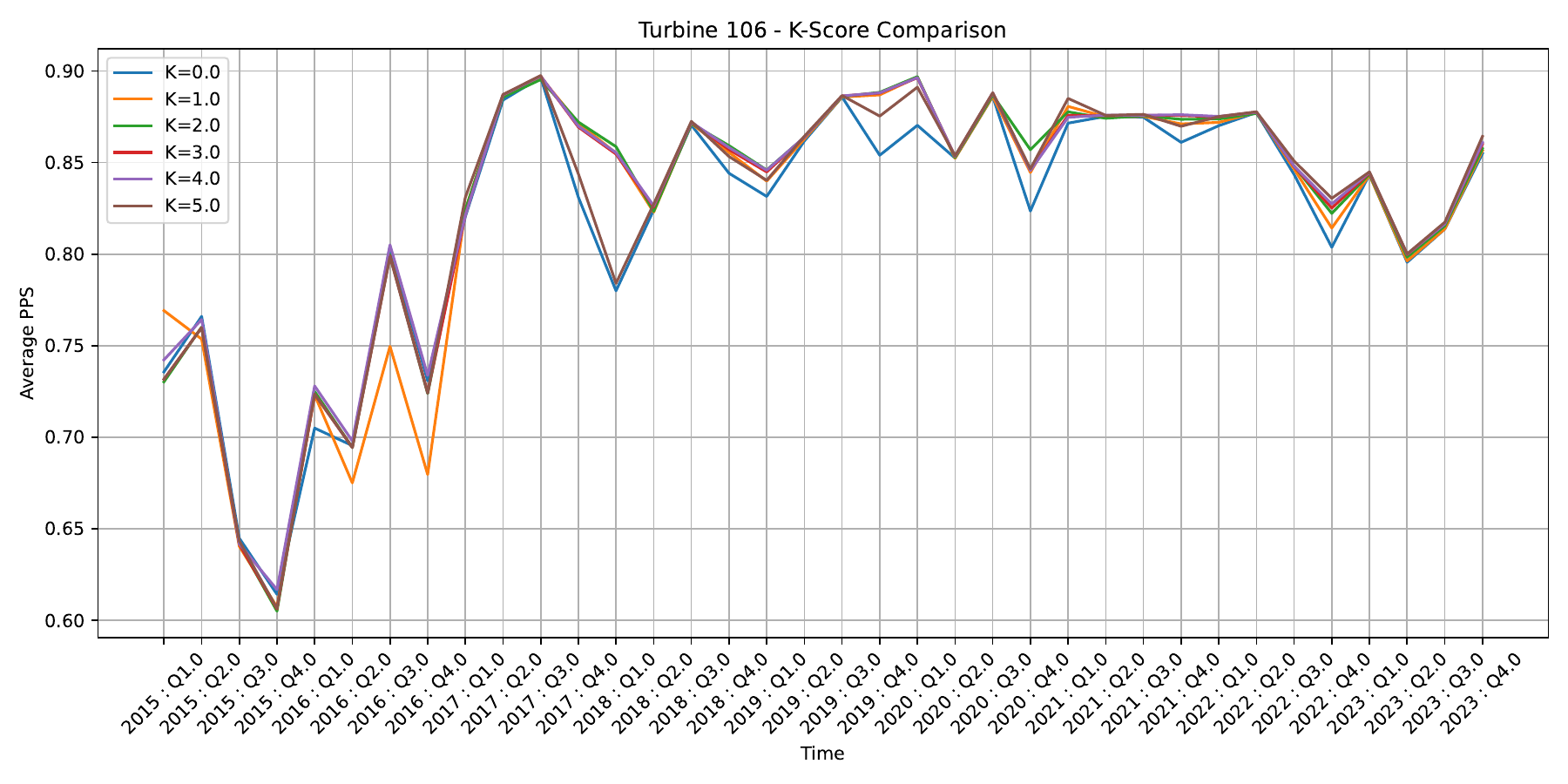}
    \caption{}
    \label{fig:apx-kgmm}
\end{figure}

\Cref{fig:apx-kgmm} presents a comparative analysis of the PPS over time for wind farm 15 wind turbine 106, evaluated under varying $k$ number of Gaussian Mixture Model components, ranging from 0 to 5.

\begin{figure}[h]
    \centering
    \includegraphics[width=1\linewidth]{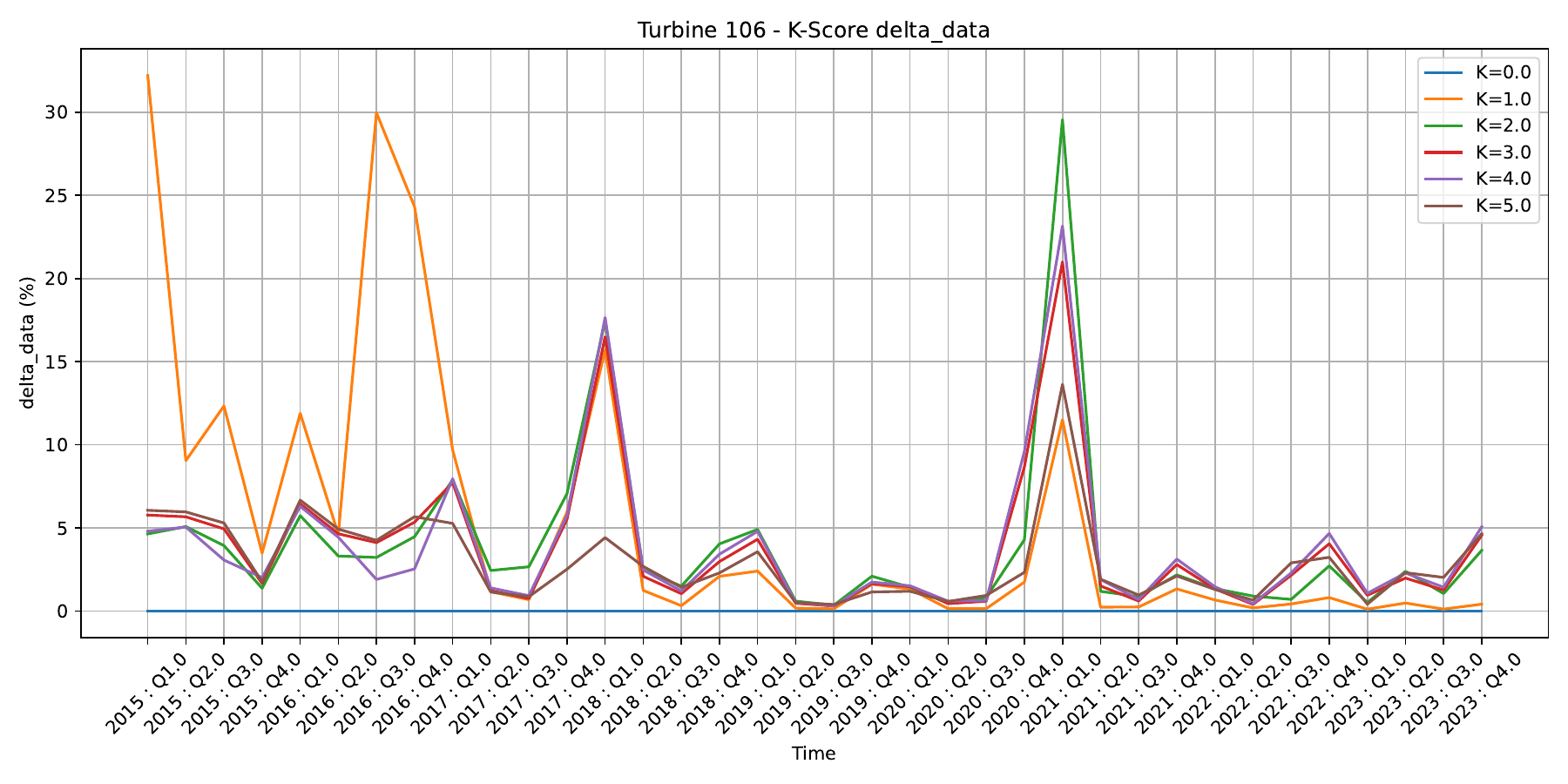}
    \caption{}
    \label{fig:apx-delta-data}
\end{figure}

\Cref{fig:apx-delta-data} presents the percentage of data removed for each $k$ in each time period.

\begin{figure}[h]
    \centering
    \includegraphics[width=1\linewidth]{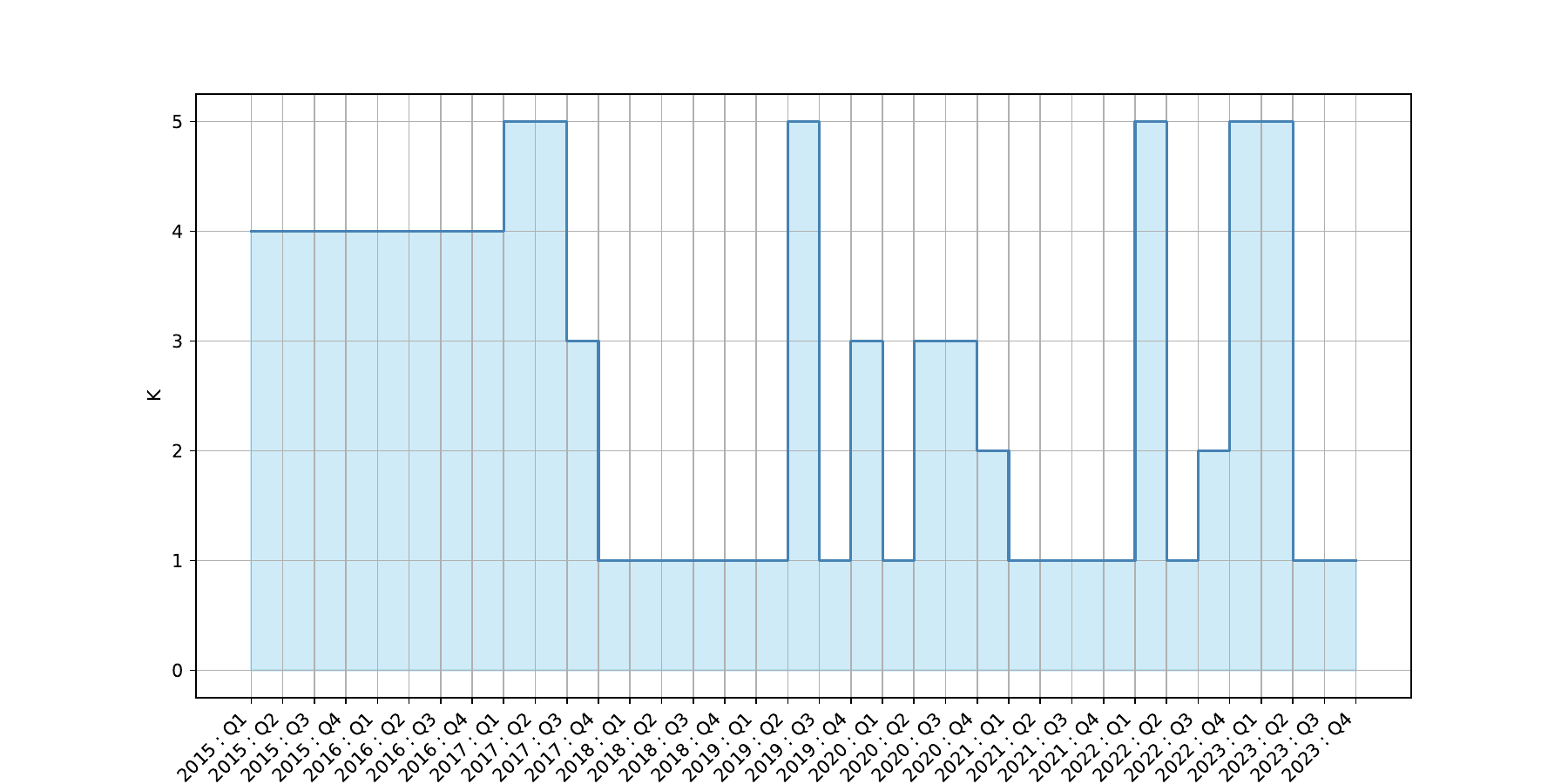}
    \caption{}
    \label{fig:apx-kselection}
\end{figure}

\Cref{fig:apx-kselection} displays the temporal evolution of the selected number of mixture components for wind farm 15 wind turbine 106 across quarterly intervals. 

\begin{figure}[h]
    \centering
    \includegraphics[width=1\linewidth]{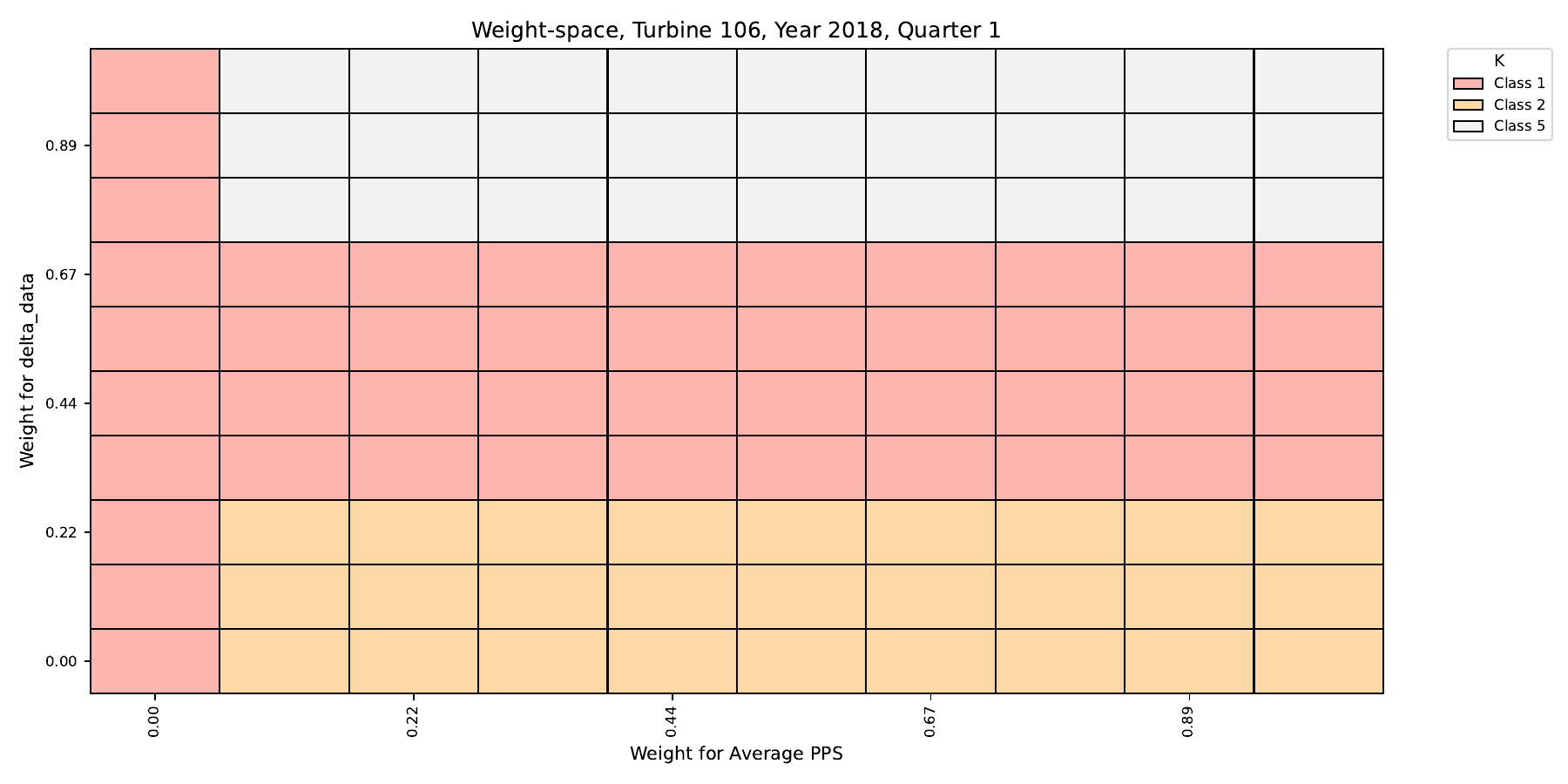}
    \caption{}
    \label{fig:apx-weights}
\end{figure}

\Cref{fig:apx-weights} explores how varying the relative weights assigned to maximizing the PPS and minimizing data removal ($N_{\delta}$) influences the chosen value of $k$. Specifically, this figure illustrates for wind farm 15 wind turbine 106 and for the time period 2018Q1. We have utilized the weights 0.6 and 0.4, respectively, for PPS and $N_{\delta}$ resulting in $k=1$.

\section{Illustration of Stable Operational Periods} \label{appendix:stable-period}
\begin{figure}[h]
    \centering
    \includegraphics[width=1\linewidth]{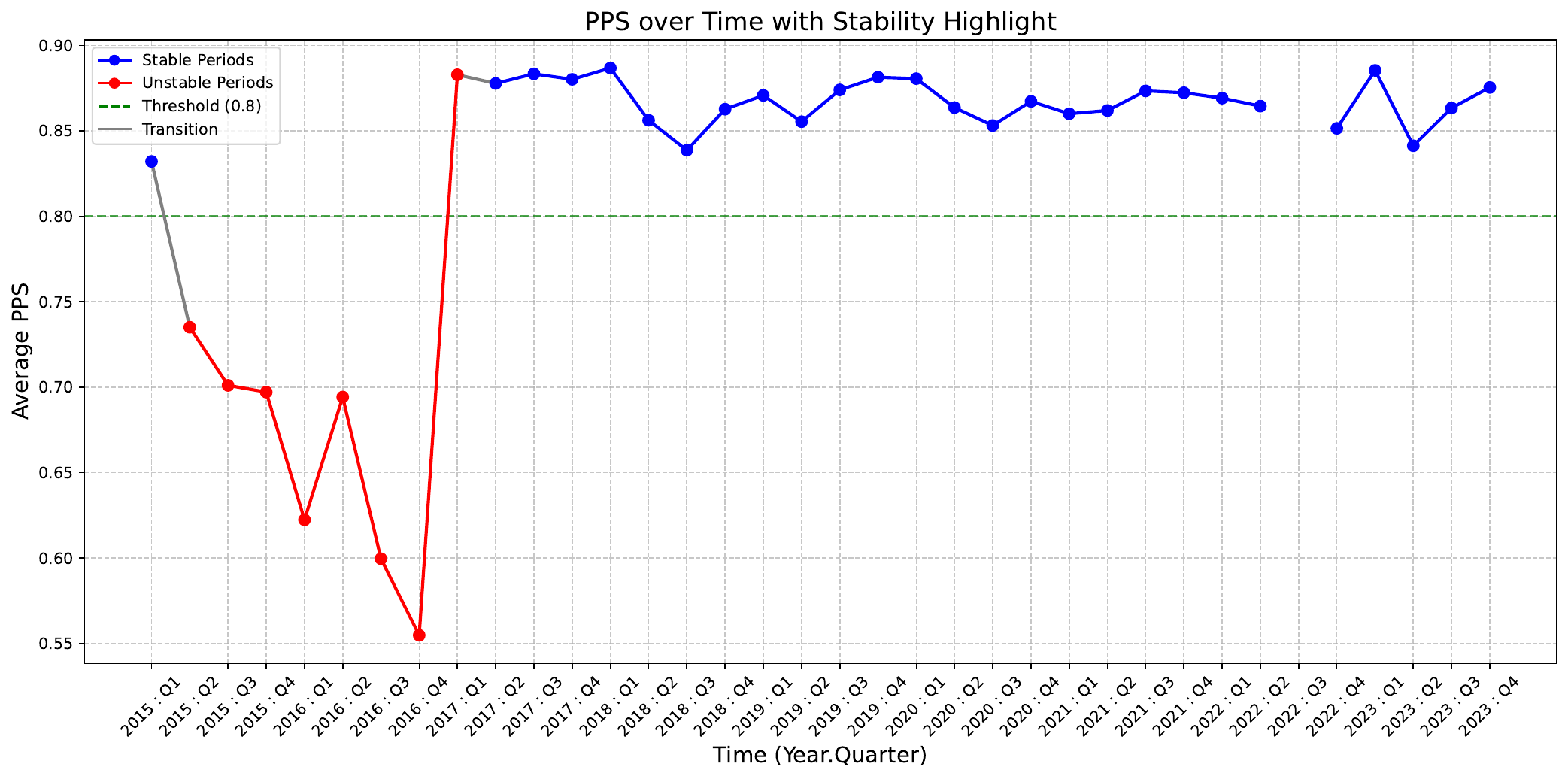}
    \caption{}
    \label{fig:apx-stability}
\end{figure}

\Cref{fig:apx-stability} illustrates the temporal evolution of the PPS for wind farm 13 wind turbine 90. The analysis incorporates two criteria to identify periods of operational stability. Firstly, the PPS must exceed a threshold value of 0.8, and secondly, the rolling standard deviation of the PPS must remain below 0.03.

\end{document}